\documentclass[useAMS,usenatbib]{mn2e}

\usepackage{amsmath}
\usepackage{graphicx}
\usepackage{txfonts}
\usepackage{natbib}

\def\aj{AJ}%
\def\araa{ARA\&A}%
\def\apj{ApJ}%
\def\apjl{ApJ}%
\def\apjs{ApJS}%
\def\aap{A\&A}%
\def\aapr{A\&A~Rev.}%
\def\aaps{A\&AS}%
\def\mnras{MNRAS}%
\def\pasa{PASA}%
\def\pasj{PASJ}%
\def\nat{Nature}%

\newcommand{\feh}{\mathrm{[Fe/H]}}
\newcommand{\teff}{T_\mathrm{eff}}
\newcommand{\logg}{\log g}
\newcommand{\fbol}{\mathcal{F}_\mathrm{bol}}
\newcommand{\ergs}{\rm{erg\,s^{-1}\,cm^{-2}\,\AA^{-1}}}

\title[The importance of appearing right]{Towards stellar effective temperatures and diameters at one per cent accuracy for future surveys}

\author[Casagrande, Portinari et al.]{\parbox{18cm}{
  L.   Casagrande$^{1}$\thanks{Stromlo Fellow}\thanks{Email:luca.casagrande@anu.edu.au}, 
  L.   Portinari$^{2}$, 
  I.S. Glass$^{3}$, 
  D.   Laney$^{4}$, 
  V.   Silva Aguirre$^{5}$, 
  J.   Datson$^{2}$,
  J.   Andersen$^{6}$, 
  B.   Nordstr\"om$^{6}$, 
  J.   Holmberg$^{6}$,
  C.   Flynn$^{7}$, 
  M.   Asplund$^{1}$}\vspace{0.3cm}\\
$^1$ Research School of Astronomy and Astrophysics, Mount Stromlo Observatory, 
        The Australian National University, ACT 2611, Australia\\
$^2$ Tuorla Observatory, Department of Physics and Astronomy, 
        University of Turku, FI-20014, Finland\\
$^3$ South African Astronomical Observatory, PO Box 9, Observatory 7935, 
        South Africa\\ 
$^4$ Bell Observatory, Western Kentucky University, Warren County KY 42274, USA\\
$^5$ Stellar Astrophysics Centre, Department of Physics and Astronomy, 
        Aarhus University, Ny Munkegade 120, DK-8000 Aarhus C, Denmark\\
$^6$ Niels Bohr Institute, Copenhagen University, Juliane Maries Vej 30, 
        2100, Copenhagen, Denmark\\ 
$^7$ Centre for Astrophysics and Supercomputing, Swinburne University of 
        Technology, VIC 3122 Australia}

\begin{document}

\date{Received; accepted}

\maketitle

\begin{abstract}
The apparent size of stars is a crucial benchmark for fundamental stellar 
properties such as effective temperatures, radii and surface gravities. While 
interferometric measurements of stellar angular diameters are the most direct 
method to gauge these, they are still limited to relatively nearby and bright 
stars, which are saturated in most of the modern photometric surveys. This 
dichotomy prevents us from safely extending well calibrated relations to the 
faint stars targeted in large spectroscopic and photometric surveys. Here, 
we alleviate this obstacle by presenting SAAO near-infrared $JHK$ observations 
of 55 stars: 16 of them have interferometric angular diameters, and the rest 
are in common with the 2MASS (unsaturated) dataset, allowing us to tie the 
effective temperatures and angular diameters derived via the Infrared Flux 
Method to the interferometric scale. We extend the test to recent 
interferometric measurements of unsaturated 2MASS stars, including giants, and 
the metal-poor benchmark target HD122563. With a critical evaluation of the 
systematics involved, we conclude that a $1$ per cent accuracy in 
fundamental stellar parameters is usually within reach.
Caution, however, must be used when indirectly testing a $\teff$ scale via 
colour relations, as well as when assessing the reliability of interferometric 
measurements, especially at sub-milliarcsec level. As a result, rather 
different effective temperature scales can be compatible with a given subset 
of interferometric data. We highlight some caveats to be aware of in such a 
quest, and suggest a simple method to check against systematics in fundamental 
measurements. A new diagnostic combination seismic radii with astrometric 
distances is also presented.
\end{abstract}


\section{Introduction}\label{sect:introduction}

Stellar angular diameters are crucial to correctly characterizing the basic 
properties of stars. Firstly, the effective temperature $\teff$ is defined as 
the temperature of a black body with the same luminosity per unit surface as 
the star. Thus, if the angular diameter is measured, the determination 
of $\teff$ only requires the additional knowledge of the bolometric flux.
Secondly, if the parallax is sufficiently well known, then it is possible 
to derive the star's intrinsic radius, competing with 
the radii obtained from eclipsing binaries to constrain stellar models 
\cite[e.g.,][and references therein]{and91,tor10}.

Interferometric measurements of stellar angular diameters have a long 
history \citep{mp21,p31,bt58}, although it was only with \cite{hb74} that 
an extensive survey was carried out and used to calibrate empirically 
the effective temperature scale for early type stars \citep{code76}.
This campaign has continued throughout the years with increasingly 
sophisticated instrumentations both in the optical and infrared (e.g.,~SUSI:
\citealt{dt86,davis11};  Mark III: \citealt{mozu91}; NPOI: \citealt{nordgren99};
IOTA: \citealt{dyck96}; PTI: \citealt{colavita99}). Recently, baselines 
exceeding $200$ and $300$ meters have been achieved on VLTI and CHARA 
respectively, yielding measurements of angular diameters for dwarfs and 
subgiants to a precision of $\sim1$ per cent or better 
\citep[e.g.,][]{kervella03,huber12,w13}. 
Although the accuracy might be lower, the aforementioned uncertainty 
translates to about $30$~K at solar $\teff$.

Arguably, the second most direct method to determine effective temperatures 
after interferometry is the InfraRed Flux Method (IRFM), a photometric 
technique originally devised to indirectly obtain angular diameters to a 
precision of few per cent, and to compete with intensity interferometry should 
a good flux calibration be achieved \citep{blackwell77,blackwell79,blackwell80}.
\cite{c06,c10} have updated the IRFM temperature scale, taking full advantage 
of the homogeneous near-infrared photometry of the 2 Micron All Sky 
Survey (2MASS), and of the extant
accuracy in the photometric zero-points and absolute flux calibration 
\citep{cohen03,bohlin07}. 
The new IRFM scale is in agreement with various spectroscopic ones 
\citep[especially for solar type stars, e.g.,][]{valenti05,fb08,sousa11}, 
but about 100~K hotter than a number of older photometric scales 
\citep[e.g.,][]{alonso96:teff_scale,rm05b} 
including the one adopted in the Geneva--Copenhagen Survey, currently the 
largest and most complete census of long-lived stars in the Solar 
Neighbourhood \citep{nordstrom04,holmberg07,holmberg09}.

Far from being a technicality, a systematic shift of $+100$~K in
effective temperature implies a shift of about $+0.1$~dex on spectroscopically 
derived metallicities \citep[e.g.,][]{melendez10}.
This shifts the peak of the 
metallicity distribution function in the Solar Neighbourhood from the 
historically accepted value of about $-0.1$~dex to roughly solar metallicity
\citep{c11}, with a number of consequences for Galactic chemical 
evolution models as well as for interpreting the Sun in a Galactic context 
\citep[e.g.,][]{wfd96,pagel:book,matteucci:book,asplund09}. A sound setting of 
the $\teff$ scale is crucial also for other reasons, e.g.~in comparison 
with theoretical stellar models or to derive absolute abundances.

The zero-point of the \cite{c10} IRFM scale was secured using solar twins, 
i.e.~stars spectroscopically selected to be virtually identical to the Sun
\citep{melendez07,melendez09:twins,ramirez09}; other, independent 
spectroscopic analyses of candidate solar twins also favour the hotter $\teff$ 
scale \citep{dfp12,datson2013,k05}.
The new scale has been tested in a number of studies, and with 
different approaches: with the colours of the Sun as derived from solar like 
stars with the model independent line--depth--ratio technique \citep{r12,c12}; 
matching the theoretical solar isochrone to the open cluster M67 
\citep{vandenberg10,pin12}; with {\it Hubble Space Telescope}
(HST) absolute spectrophotometry in the metal-poor regime \citep{c10}.

However the key test remains the comparison to angular diameters from 
direct interferometric measurements. So far, this was impeded by virtually 
zero overlap between stars with high precision interferometric data 
and (unsaturated) 2MASS photometry, which is at the base of any modern 
implementation of the IRFM.
In \cite{c10}, an indirect comparison was performed by deriving angular 
diameters via optical colour--metallicity--$\teff$ and 
colour--magnitude--metallicity--bolometric flux relations, and the comparison 
to interferometry proved excellent. Although the optical relations were 
calibrated on the sample stars defining the IRFM scale, the comparison did not 
rely {\it directly} on the IRFM \citep[see also discussion in][]{c08:uppsala}. 

Here, we overcome this major limitation and perform a direct, potentially 
conclusive test running the IRFM on 16 bright stars having angular diameter 
measurements. The trivial option of transforming existing 
near-infrared photometry of nearby stars into the 2MASS system is sub-optimal 
(see Section~\ref{sect:IRFM}) and thus we resort to new dedicated $JHK$ 
photometry of nearby stars measured at the South African Astronomical 
Observatory (SAAO).
The two IRFM scales (SAAO and 2MASS--based) are tied together via 38 stars in 
common between the two systems.

While in this work we concentrate on stars with interferometric diameters 
above and around 1~mas,
the increasing capabilities of CHARA \citep[especially with 
the PAVO beam combiner,][]{chara_pavo} and 
repeated, careful observations are now pushing the limit for reliable angular 
diameters down to $\sim 0.5$~mas \citep[][]{w13}. This finally allows us to 
target stars having good 2MASS photometry and directly test a number of 
effective temperature scales as well as other interferometric measurements 
\citep{huber12}. 
Thus, this is the first time the 2MASS effective temperature and absolute flux 
scale is tested with high precision interferometric angular diameters. 

The plan of the paper is as follows. In Section~\ref{sect:observations} we 
present the near-infrared photometry measured from SAAO. 
In Section~\ref{sect:IRFM} we briefly recall the basic principles of the IRFM,
and derive the fundamental parameters of the sample stars, which we test 
against interferometric measurements in Section~\ref{sect:zeropoint}.
In Section~\ref{sect:GCS} we discuss the zero-point of the Geneva-Copenhagen 
Survey scale and of interferometric data. A simple 
exercise to highlight systematics and test the internal consistency of 
fundamental measurements is laid out. As it is now possible  to estimate 
stellar radii from asteroseismic scaling relations, in 
Section~\ref{sect:parallaxes} we present a new approach to gauge the angular 
diameter and effective temperature scale by coupling parallaxes with 
asteroseismology. In Section~\ref{sect:Conclusions} we draw our conclusions.

\section{Near-infrared SAAO observations}
\label{sect:observations}

The SAAO $JHK$ photometric system was established by \cite{g74} and its 
accuracy and zero-points refined and improved over the years by \cite{c90}
and \cite{cm95}. Photometric observations were carried using the MkII IRP 
on the SAAO $0.75$~m telescope, typically with a diaphragm size of 36 arcsec 
and with a fixed chopping amplitude of about $180$ arcsec on a north-south 
line. Standard stars from the Carter list were observed roughly every hour.

SAAO photometry for the full sample is reported in Table~\ref{t:saao}.
Around two thirds of the stars have double or multiple observations, 
and for them we list average magnitudes and corresponding $1~\sigma$ scatter. 
Based on the latter, we estimate typical photometric errors to be within 
$0.01-0.02$~mag in all three bands \citep[consistent with][]{c90}, 
and assume this value for stars having only one measurement.

For our sample stars having also accurate 2MASS photometry, the average 
difference between the two systems (2MASS-SAAO) is $-0.047 \pm 0.023$~mag
in $J$, $-0.011 \pm 0.024$~mag in $H$ and $-0.032 \pm 0.016$~mag in $K$.
These are, within the small colour range covered by our sample, in very good 
agreement with the updated transformations of \cite{carpenter01}\footnote{http://www.astro.caltech.edu/$\sim$jmc/2mass/v3/transformations}. 
Notice though, that no transformation between any photometric systems is 
done in this work: for the sake of the IRFM is in fact more robust to work 
directly with physical quantities (i.e.~to implement the proper filter 
transmission curves, zero points and absolute fluxes to translate magnitudes 
into fluxes) rather than converting magnitudes between different photometric 
systems, as explained later in more detail.

\begin{table*}
\centering
\caption{Photometry and adopted parameters of sample stars}\label{t:saao}
\begin{tabular}{cclll|ccccc|cr}
\hline \hline
     HD  &    HIP  &      ~~~~$J$       &      ~~~~$H$       &    ~~~~$K$         &    $V$ &  $B-V$ & $V-R_C$& $V-I_C$&   Ref  &  $\logg$    & $\feh$   \\   
\hline 
   30652 &   22449 &  2.380 &  2.130 &  2.104 & 3.190 &  0.460 &  0.270 &  0.525 &    B90 &  $4.29^{c}$ & $-0.02^{a}$  \\
   38973 &   27244 &  5.615 &  5.304 &  5.259 &  $-$  &   $-$  &   $-$  &   $-$  &   $-$  &  $4.39^{a}$ & $ 0.00^{a}$  \\
   39587 &   27913 &  3.387 &  3.035 &  2.988 &  $-$  &   $-$  &   $-$  &   $-$  &   $-$  &  $4.47^{a}$ & $-0.12^{a}$  \\
   48737 &   32362 &  $2.574 \pm 0.003$ &  $2.323 \pm 0.002$ &  $2.291 \pm 0.002$ &  $-$  &   $-$  &   $-$  &   $-$  &   $-$  &  $3.83^{a}$ & $ 0.19^{a}$  \\
   52298 &   33495 &  6.003 &  5.748 &  5.695 &  $-$  &   $-$  &   $-$  &   $-$  &   $-$  &  $4.37^{a}$ & $-0.30^{a}$  \\
   56537 &   35350 &  $3.390 \pm 0.011$ &  $3.309 \pm 0.021$ &  $3.262 \pm 0.071$ &  $-$  &   $-$  &   $-$  &   $-$  &   $-$  &  $3.90^{d}$ & $-0.10^{d}$  \\
   58192 &   35884 &  5.991 &  5.712 &  5.674 &  $-$  &   $-$  &   $-$  &   $-$  &   $-$  &  $4.38^{a}$ & $-0.22^{a}$  \\
   69655 &   40438 &  5.575 & 5.265 &  5.219 &  $-$  &   $-$  &   $-$  &   $-$  &   $-$  &  $4.40^{a}$ & $-0.20^{a}$  \\
   71334 &   41317 &  $6.670 \pm 0.007$ &  $6.301 \pm 0.018$ &  $6.251 \pm 0.003$ & 7.809 &  0.664 &  0.367 &  0.714 &    R12 &  $4.45^{a}$ & $-0.05^{a}$  \\
   75289 &   43177 &  5.375 &  5.108 &  5.054 &  $-$  &   $-$  &   $-$  &   $-$  &   $-$  &  $4.32^{a}$ & $ 0.21^{a}$  \\
   75732 &   43587 &  $4.576 \pm 0.015$ &  $4.132 \pm 0.003$ &  $4.069 \pm 0.006$ &  $-$  &   $-$  &   $-$  &   $-$  &   $-$  &  $4.45^{c}$ & $ 0.31^{c}$  \\
   76151 &   43726 &  $4.894 \pm 0.022$ &  $4.553 \pm 0.011$ &  $4.499 \pm 0.005$ & 6.000 &  0.670 &  0.360 &  0.695 &    B90 &  $4.43^{a}$ & $ 0.03^{a}$  \\
   78534 &   44935 &  $7.591 \pm 0.013$ &  $7.236 \pm 0.011$ &  $7.188 \pm 0.015$ & 8.688 &  0.654 &  0.345 &  0.684 &    R12 &  $4.41^{b}$ & $ 0.07^{b}$  \\
   78660 &   44997 &  7.196 &  6.843 &  6.787 & 8.325 &  0.666 &  0.344 &  0.685 &    R12 &  $4.39^{a}$ & $-0.03^{a}$  \\
   82943 &   47007 &  5.517 &  5.204 &  5.156 &  $-$  &   $-$  &   $-$  &   $-$  &   $-$  &  $4.38^{a}$ & $ 0.30^{a}$  \\
   83683 &   47468 &  6.056 &  5.777 &  5.755 &  $-$  &   $-$  &   $-$  &   $-$  &   $-$  &  $4.29^{a}$ & $-0.18^{a}$  \\
   86226 &   48739 &  $6.860 \pm 0.013$ &  $6.548 \pm 0.011$ &  $6.489 \pm 0.001$ &  $-$  &   $-$  &   $-$  &   $-$  &   $-$  &  $4.45^{a}$ & $ 0.00^{a}$  \\
   87359 &   49350 &  $6.293 \pm 0.014$ &  $5.927 \pm 0.012$ &  $5.859 \pm 0.011$ &  $-$  &   $-$  &   $-$  &   $-$  &   $-$  &  $4.42^{a}$ & $ 0.05^{a}$  \\
   88072 &   49756 &  $6.432 \pm 0.002$ &  $6.110 \pm 0.010$ &  $6.049 \pm 0.001$ & 7.525 &  0.644 &  0.349 &  0.672 &    R12 &  $4.45^{a}$ & $ 0.01^{a}$  \\
   91638 &   51784 &  5.725 &  5.439 &  5.397 &  $-$  &   $-$  &   $-$  &   $-$  &   $-$  &  $4.30^{a}$ & $-0.20^{a}$  \\
   92719 &   52369 &  5.674 &  5.362 &  5.286 &  $-$  &   $-$  &   $-$  &   $-$  &   $-$  &  $4.48^{a}$ & $-0.15^{a}$  \\
   93372 &   52535 &  5.397 &  5.148 &  5.112 &  $-$  &   $-$  &   $-$  &   $-$  &   $-$  &  $4.31^{a}$ & $ 0.05^{a}$  \\
   94690 &   53424 &  $7.056 \pm 0.012$ &  $6.697 \pm 0.011$ &  $6.622 \pm 0.012$ &  $-$  &   $-$  &   $-$  &   $-$  &   $-$  &  $4.35^{a}$ & $ 0.25^{a}$  \\
   96700 &   54400 &  $5.451 \pm 0.026$ &  $5.101 \pm 0.023$ &  $5.057 \pm 0.021$ & 6.530 &  0.600 &  0.340 &  0.670 &    C80 &  $4.32^{a}$ & $-0.29^{a}$  \\
   97603 &   54872 &  2.277 &  2.229 &  2.201 &  $-$  &   $-$  &   $-$  &   $-$  &   $-$  &  $3.90^{d}$ & $ 0.06^{d}$  \\
  101805 &   57092 &  $5.560 \pm 0.002$ &  $5.297 \pm 0.002$ &  $5.245 \pm 0.006$ & 6.471 &  0.524 &  0.295 &  0.573 &    M89 &  $4.29^{a}$ & $ 0.11^{a}$  \\
  102870 &   57757 &  $2.641 \pm 0.010$ &  $2.356 \pm 0.007$ &  $2.308 \pm 0.004$ & 3.600 &  0.550 &  0.320 &  0.610 &    B90 &  $4.22^{c}$ & $ 0.21^{a}$  \\
  103975 &   58380 &  $5.814 \pm 0.006$ &  $5.524 \pm 0.007$ &  $5.483 \pm 0.002$ & 6.766 &  0.522 &  0.298 &  0.592 &    M89 &  $4.30^{a}$ & $-0.03^{a}$  \\
  107692 &   60370 &  $5.631 \pm 0.028$ &  $5.302 \pm 0.001$ &  $5.245 \pm 0.010$ & 6.703 &  0.651 &  0.349 &  0.674 &    R12 &  $4.44^{a}$ & $ 0.20^{a}$  \\
  114174 &   64150 &  $5.670 \pm 0.018$ &  $5.299 \pm 0.004$ &  $5.249 \pm 0.005$ & 6.761 &  0.688 &  0.349 &  0.694 &    R12 &  $4.38^{a}$ & $ 0.05^{a}$  \\
  114853 &   64550 &  $5.774 \pm 0.001$ &  $5.411 \pm 0.002$ &  $5.361 \pm 0.010$ &  $-$  &   $-$  &   $-$  &   $-$  &   $-$  &  $4.46^{a}$ & $-0.16^{a}$  \\
  115169 &   64713 &  $8.111 \pm 0.005$ &  $7.769 \pm 0.005$ &  $7.720 \pm 0.017$ & 9.250 &  0.648 &  0.355 &  0.690 &    R12 &  $4.52^{b}$ & $-0.01^{b}$  \\
  118098 &   66249 &  $3.161 \pm 0.026$ &  $3.096 \pm 0.021$ &  $3.077 \pm 0.021$ & 3.380 &  0.110 &  0.062 &  0.122 &    C80 &  $4.02^{d}$ & $ 0.16^{d}$  \\
  121560 &   68030 &  $5.196 \pm 0.003$ &  $4.894 \pm 0.003$ &  $4.858 \pm 0.016$ &  $-$  &   $-$  &   $-$  &   $-$  &   $-$  &  $4.34^{a}$ & $-0.30^{a}$  \\
  131977 &   73184 &  $3.841 \pm 0.003$ &  $3.240 \pm 0.007$ &  $3.146 \pm 0.003$ & 5.760 &  1.060 &  0.650 &  1.180 &    C86 &  $4.76^{c}$ & $ 0.31^{a}$  \\
  132301 &   73383 &  $5.718 \pm 0.003$ &  $5.483 \pm 0.002$ &  $5.453 \pm 0.003$ & 6.582 &  0.471 &  0.276 &  0.540 &    M89 &  $4.34^{a}$ & $-0.03^{a}$  \\
  138573 &   76114 &  $6.102 \pm 0.009$ &  $5.742 \pm 0.010$ &  $5.683 \pm 0.006$ &  $-$  &   $-$  &   $-$  &   $-$  &   $-$  &  $4.41^{a}$ & $-0.04^{a}$  \\
  141795 &   77622 &  $3.489 \pm 0.015$ &  $3.437 \pm 0.001$ &  $3.420 \pm 0.003$ & 3.710 &  0.150 &  0.065 &  0.129 &    C80 &  $4.24^{d}$ & $ 0.23^{d}$  \\
  142331 &   77883 &  $7.556 \pm 0.010$ &  $7.194 \pm 0.010$ &  $7.149 \pm 0.012$ & 8.727 &  0.681 &  0.368 &  0.719 &    R12 &  $4.39^{b}$ & $ 0.04^{b}$  \\
  142860 &   78072 &  $2.961 \pm 0.023$ &  $2.685 \pm 0.024$ &  $2.647 \pm 0.013$ &  $-$  &   $-$  &   $-$  &   $-$  &   $-$  &  $4.18^{c}$ & $-0.14^{a}$  \\
  143436 &   78399 &  $6.943 \pm 0.004$ &  $6.609 \pm 0.027$ &  $6.564 \pm 0.018$ &  $-$  &   $-$  &   $-$  &   $-$  &   $-$  &  $4.28^{f}$ & $ 0.00^{f}$  \\
  145825 &   79578 &  5.436 &  5.095 &  5.058 & 6.533 &  0.678 &  0.352 &  0.699 &    R12 &  $4.47^{a}$ & $ 0.12^{a}$  \\
  146233 &   79672 &  $4.391 \pm 0.016$ &  $4.057 \pm 0.016$ &  $4.006 \pm 0.015$ & 5.510 &  0.650 &  0.357 &  0.691 &    R12 &  $4.45^{b}$ & $ 0.05^{b}$  \\
  153458 &   83181 &  $6.880 \pm 0.005$ &  $6.538 \pm 0.012$ &  $6.483 \pm 0.009$ &  $-$  &   $-$  &   $-$  &   $-$  &   $-$  &  $4.42^{a}$ & $ 0.07^{a}$  \\
  157338 &   85158 &  $5.878 \pm 0.032$ &  $5.565 \pm 0.005$ &  $5.508 \pm 0.005$ &  $-$  &   $-$  &   $-$  &   $-$  &   $-$  &  $4.36^{a}$ & $-0.17^{a}$  \\
  157347 &   85042 &  $5.118 \pm 0.005$ &  $4.760 \pm 0.012$ &  $4.708 \pm 0.005$ & 6.287 &  0.669 &  0.364 &  0.707 &    R12 &  $4.42^{a}$ & $ 0.03^{a}$  \\
  159063 &   85799 &  6.095 &  5.834 &  5.800 &  $-$  &   $-$  &   $-$  &   $-$  &   $-$  &  $4.26^{a}$ & $ 0.22^{a}$  \\
  164259 &   88175 &  $3.921 \pm 0.010$ &  $3.718 \pm 0.015$ &  $3.691 \pm 0.013$ & 4.620 &  0.390 &  0.227 &  0.452 &    B90 &  $4.08^{a}$ & $-0.08^{a}$  \\
  167060 &   89650 &  $7.842 \pm 0.004$ &  $7.504 \pm 0.003$ &  $7.458 \pm 0.008$ & 8.943 &  0.644 &  0.354 &  0.679 &    R12 &  $4.48^{b}$ & $ 0.02^{b}$  \\
  173667 &   92043 &  3.352 &  3.108 &  3.062 &  $-$  &   $-$  &   $-$  &   $-$  &   $-$  &  $3.98^{c}$ & $-0.01^{a}$  \\
  177724 &   93747 &  $2.938 \pm 0.012$ &  $2.917 \pm 0.004$ &  $2.880 \pm 0.014$ &  $-$  &   $-$  &   $-$  &   $-$  &   $-$  &  $3.74^{e}$ & $-0.10^{e}$  \\
  182572 &   95447 &  $3.958 \pm 0.023$ &  $3.595 \pm 0.028$ &  $3.504 \pm 0.002$ & 5.143 &  0.769 &  0.381 &  0.740 &    R12 &  $4.32^{c}$ & $ 0.40^{c}$  \\
  184509 &   96370 &  $5.760 \pm 0.014$ &  $5.446 \pm 0.024$ &  $5.400 \pm 0.014$ &  $-$  &   $-$  &   $-$  &   $-$  &   $-$  &  $4.32^{a}$ & $-0.19^{a}$  \\
  189931 &   98813 &  $5.839 \pm 0.006$ &  $5.511 \pm 0.015$ &  $5.458 \pm 0.013$ &  $-$  &   $-$  &   $-$  &   $-$  &   $-$  &  $4.45^{a}$ & $ 0.01^{a}$  \\
  194640 &  100925 &  $5.390 \pm 0.010$ &  $5.009 \pm 0.012$ &  $4.945 \pm 0.012$ & 6.615 &  0.730 &  0.392 &  0.762 &    C80 &  $4.48^{a}$ & $-0.01^{a}$  \\
\hline
\end{tabular}
\begin{list}{}{}
\item[] 
  Source of Johnson-Cousins photometry: \cite{c80}, 
  \cite{c86}, \cite{m89}, \cite{b90}, \cite{r12}. Adopted stellar parameters:
       {\it a-} \cite{c11}; {\it b-} \cite{r12}; {\it c-} \cite{valenti05}; 
       {\it d-} \cite{tak09}; {\it e-} \cite{gc03}; {\it f-} \cite{k05}.
\end{list}
\end{table*}

\section{Fundamental parameters from the InfraRed Flux Method}\label{sect:IRFM}

The IRFM, an elegant and almost model--independent photometric technique
for determining angular diameters and effective temperatures, has been 
implemented by 
various authors over the years \citep[e.g.,][]{blackwell77, blackwell79, 
blackwell80, bl94, bell89, alonso96:irfm, rm05a, ghb09, c06, c10}. 
While we refer to the aforementioned papers for a detailed description of the 
method, we briefly recall the key points relevant for the present work.

The IRFM relies on the ratio between the bolometric flux ($\fbol$) and the 
infrared monochromatic 
flux ($\mathcal{F}_\mathrm{IR}$) of a star measured on the Earth. This 
ratio is compared to the one defined on a stellar surface element as follows:
\begin{equation}\label{eq:irfm}
\frac{\fbol\rm{(Earth)}}{\mathcal{F}_\mathrm{IR}\rm{(Earth)}} = 
\frac{\sigma \teff^4}{\mathcal{F}_\mathrm{IR}\rm{(model)}}.
\end{equation}
Since $\teff$ is the only 
unknown quantity, it can be readily obtained. The crucial advantage of this 
procedure over other photometric techniques is that, 
at least for spectral types earlier than $\sim\rm{M}0$, near-infrared 
photometry of stars samples the Rayleigh--Jeans tail of their spectrum, a 
region largely dominated by the continuum \citep[but see][for a discussion of 
the importance of $\rm{H}^{-}$ opacity]{blp91}, with a roughly linear 
dependence on $\teff$ and very little affected by other stellar parameters 
such as metallicity and surface gravity 
\citep[as extensively tested in literature, e.g.,][]{alonso96:irfm,
c06,c10} and nearly free from non-LTE and granulation effects 
\citep[][]{asplund01,c09}. The method (Eq.~\ref{eq:irfm}) yields 
{\it self-consistently} the effective temperature and bolometric flux of 
a star, from which its angular diameter ($\theta$) can be trivially derived 
$\fbol\rm{(Earth)}=(\theta/2)^2 \sigma$$\teff^4$.
Since most of the times multi-band photometry is used, the problem is 
ultimately reduced to a proper derivation of physical fluxes ($\ergs$) from 
magnitudes, i.e.\ to the underlying photometric absolute calibration.

Without exaggeration, this is the most critical point when implementing the 
IRFM, as already recognized in \cite{bp90}. \cite{c10} further highlighted 
how any difference between IRFM scales in the literature could be simply 
explained by changing the 
absolute calibration of the adopted photometric systems, or equivalently using 
different photometric zero-points. In this sense, changing filter sets and/or 
zero-points corresponds to introducing different IRFM scales. This also 
implies that homogeneous and well standardized 
photometry must be used, and filter transformation from one system to the 
other preferentially avoided, since systematic zero-point offsets are often 
hidden in the scatter of different colour transformations. This was the main 
motivation to obtain and analyze in this work dedicated near-infrared 
photometry for our sample stars.

\subsection{The IRFM in this work}

We use the same IRFM implementation described in \citet[][and references 
therein]{c06,c10}, where the relevant formalism on transforming heterochromatic 
measurements into monochromatic quantities at the corresponding star plus 
filter effective wavelength, taking into account energy or photo-counting 
integration, can also be found. 
The bolometric flux is recovered using multi-band optical and 
near-infrared photometry and the flux outside of these bands is estimated 
using a theoretical model flux at a given $\teff$, $\feh$ and $\logg$. The 
adopted $\feh$ and $\logg$ for each star are reported in Table~\ref{t:saao},
while an iterative procedure is adopted to converge in $\teff$.
For internal consistency,
we preferred gravity and metallicity data from \cite{c11}, or from 
spectroscopic studies adopting a $\teff$ scale consistent with theirs; but the
specific choice of $\feh$, $\logg$ and model atmospheres typically affects the 
IRFM temperatures, separately, by $\sim 10$~K at most, for the reasons 
explained above \citep[see similar comparison e.g.~in][]{alonso96:irfm}.

The effect of random photometric errors on $\fbol, \teff$ and $\theta$ for 
each star are derived using a Monte Carlo simulation and added in quadrature 
to the uncertainty stemming from a change of $\pm 0.5$~dex in $\logg$ and 
$\pm 0.2$~dex in metallicity. The error in metallicity includes a typical
$0.1$~dex precision of abundance determinations, and an 
additional systematic uncertainty by the same amount, corresponding to 
a possible shift of $100$~K in the assumed $\teff$ --- which is the 
accuracy we aim to test. (Notice though, that such metallicity--temperature
interplay only refers to spectroscopic estimates; in the IRFM, a change of 
$0.1$~dex in $\feh$ affects the temperatures typically by less than $10$~K.)
Finally, we increased all errors by an additional $20$~K in effective 
temperature, $1.0$ per cent in $\fbol$ and $0.7$ per cent in $\theta$, which 
are the zero-point uncertainties derived in \cite{c10}.

All of our sample stars have {\it Hipparcos} distances closer than $72$~pc 
\citep{vanLeeuwen07}, and are well within the local bubble, where reddening is 
negligible \citep[e.g.,][]{leroy93,lallement13}. 
This is important for robust IRFM results, since a change of $0.01$~mag 
in $E(B-V)$ would affect the $\teff$ at the level of $50$~K \citep{c10}.

In the following subsections we present the results obtained implementing the 
IRFM in different photometric systems (i.e.~with the filter transmission 
curves, zero-points and absolute fluxes appropriate to each), which 
effectively correspond to introducing (slightly) different IRFM scales. 
In the optical we use the Tycho2 system, well standardized and homogeneous for 
magnitudes brighter than about $10$ which is always the case in the present 
study. For a subset of stars, Johnson-Cousins photometry is also 
available and used; while it provides a more complete coverage of the 
optical part of a spectrum, in the $\teff$ range explored here this choice is 
fortunately irrelevant (see Section \ref{sect:Johnson}). Also, 
until recently optical photometry of solar twins was available only in the 
Tycho2 system; this gap has now been amended thanks to the observational 
efforts of \cite{r12} and the solar calibration of the IRFM tested in both 
Tycho2 and Johnson-Cousins system \citep{c12}.
However, what really drives the derived stellar parameters in our 
technique is of course the infrared photometry, i.e.~2MASS $JHK_s$ and SAAO 
$JHK$ magnitudes. 
Since we ultimately aim to test the calibration of our IRFM scale in the 
widely used 2MASS system, the best approach is to adjust the SAAO absolute flux 
calibration to yield effective temperatures and diameters consistent with 
2MASS. Within the IRFM this is far more robust way of tying the two systems 
together, rather than a star-by-star conversion of magnitudes.
We discuss all these subtleties further below.

\subsubsection{Tycho2--2MASS}\label{sect:Tycho2mass}

The IRFM implementing the Tycho2 $B_TV_T$ \citep{hog2000} and 2MASS $JHK_S$ 
\citep{cutri03} photometric system takes advantage of the well defined 
absolute calibration in the infrared \citep{cohen03,rieke08}: its zero-point 
is calibrated with a claimed accuracy of $\sim 20$~K using solar twins 
\citep{c10}. Thirty-eight stars in Table \ref{t:saao} satisfy the 2MASS quality
requirement for reliably applying the IRFM 
(i.e.~``j\_''$+$``h\_''$+$``k\_msigcom''$<0.15$), all but one having 
photometric quality flag ``AAA''\footnote{http://www.ipac.caltech.edu/2mass/releases/allsky/doc/sec2\_2a.html}. 

\subsubsection{Tycho2--SAAO}\label{sect:SAAO-2MASS}

The SAAO $JHK$ filter set has been implemented in our IRFM procedure.
Observationally the zero-points of the SAAO $JHK$ photometric system are based 
on 25 early main-sequence stars \citep{c90}. 
Since Vega is unobservable in the Southern hemisphere, it can not be used as 
primary flux calibrator, and  Sirius is often chosen as a complementary 
or alternative standard \citep[e.g.,][]{cohen92}; \cite{c08} derived 
the absolute calibration of the SAAO system by scaling a Kurucz synthetic 
spectrum of Sirius with the interferometric angular diameter measurement of 
\cite{ker03}. However, resorting to this absolute calibration would introduce 
yet a slightly different $\teff$ scale from the 2MASS--based one we wish to 
test. 

\begin{figure*}
\begin{center}
\includegraphics[width=8.5truecm]{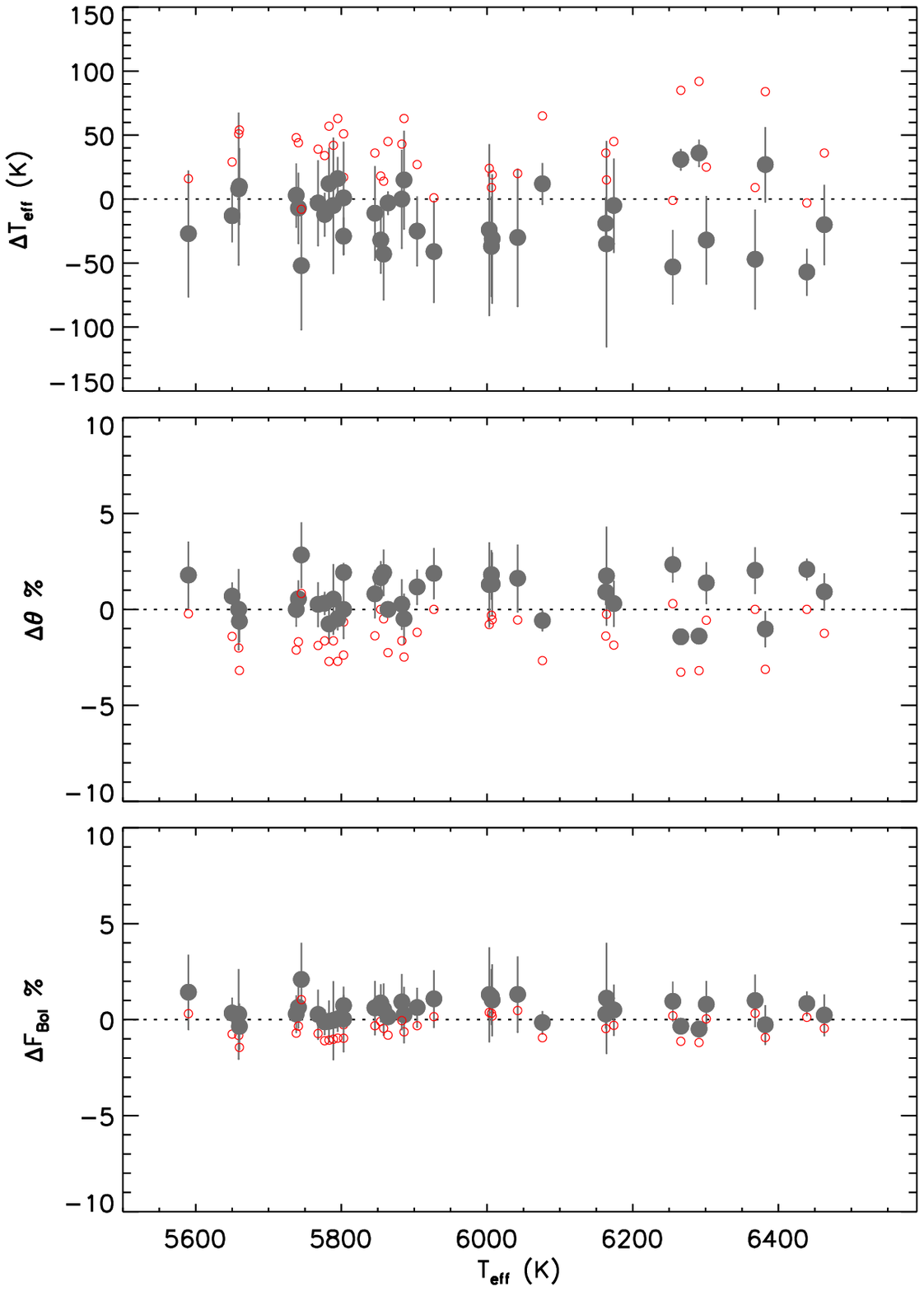}
\includegraphics[width=8.5truecm]{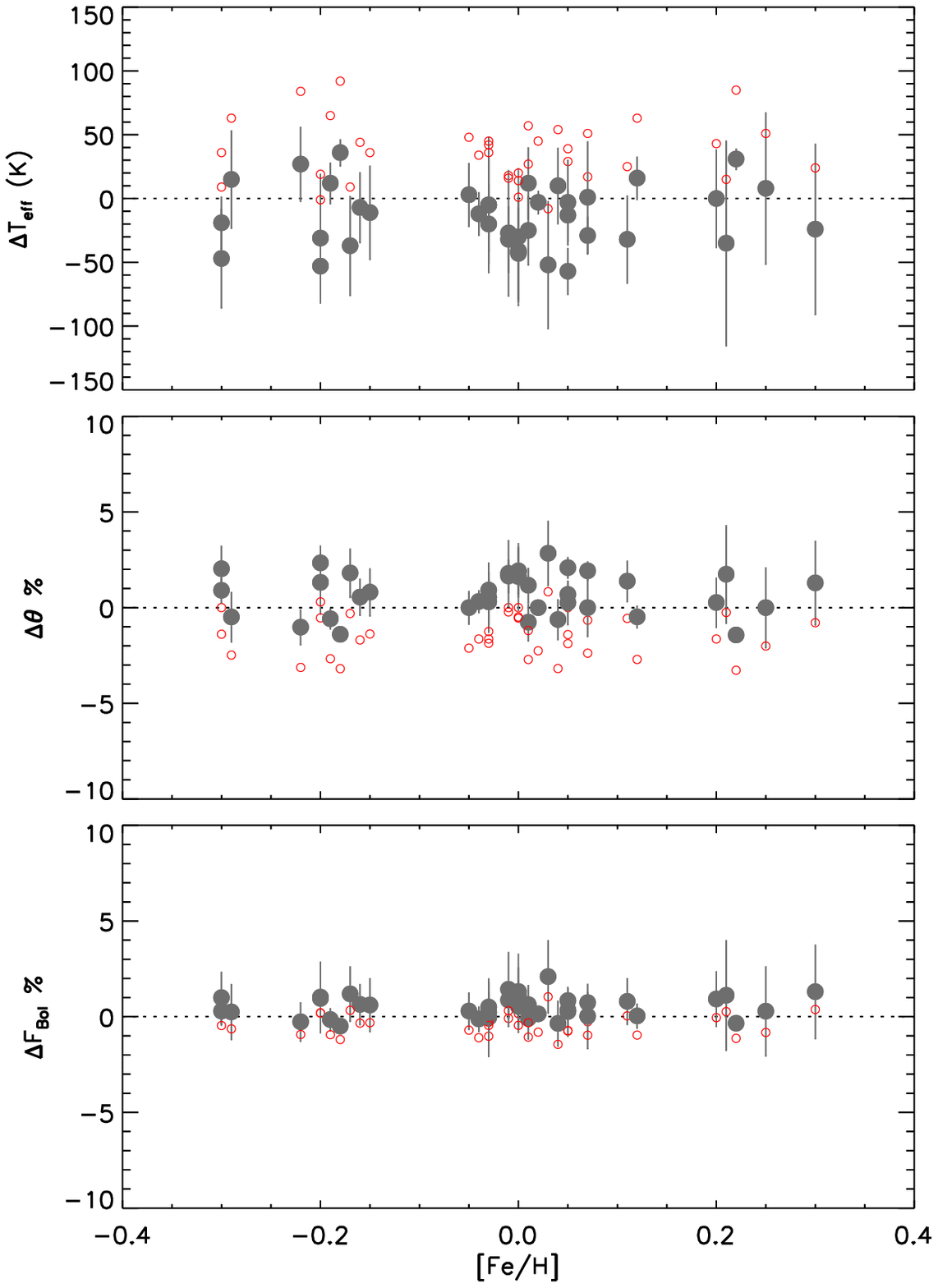}
\caption{Comparison (SAAO minus 2MASS) of $\teff, \theta$ and $\fbol$ obtained 
implementing the two systems in the IRFM, for 38 stars in common. Open circles 
refer to the comparison before adjusting the SAAO absolute flux calibration, 
filled points after increasing it by 3 per cent.}
\label{fig:2S}
\end{center}
\end{figure*}

Therefore, we opt to let the SAAO $JHK$ absolute calibration vary, until the 
resulting weighted average of $\teff, \theta$ and $\fbol$ derived from
Tycho2--SAAO photometry agree with those from Tycho2--2MASS for 38 stars in 
common (Figure \ref{fig:2S}). Apart from the infrared photometry, all other 
input parameters are the same ($B_TV_T,\feh,\logg$), thus in the weighted 
average only the internal accuracy is considered, and this can be immediately 
estimated from the scatter 
returned from each infrared band used into the IRFM. With a 3 per cent increase
in the SAAO absolute flux calibration of \cite{c08}, we achieve an agreement 
with the Tycho2--2MASS based scale of $\Delta\teff = -1 \pm 4$~K 
($\sigma=24$~K), $\Delta\theta=-0.01\pm0.11$ per cent ($\sigma=1.10$ per cent) 
and $\Delta\fbol=+0.1\pm0.1$ per cent ($\sigma=0.6$ per cent). The 3 per cent 
adjustment in absolute calibration is relatively minor, and still in accordance 
with the 2 per cent uncertainty estimated in Casagrande et al.\,(2008; we 
recall here that 
the best absolute flux scale currently available from the HST is at the per 
cent level, see \citealt{bohlin07}); without this, SAAO--based IRFM 
temperatures would be systematically hotter by about 30~K.

\subsubsection{Johnson-Cousins--SAAO}\label{sect:Johnson}

A subset of 25 stars also has Johnson-Cousins $BV(RI)_C$ photometry from 
the literature (Table~\ref{t:saao}); this allows us to implement the IRFM 
using this system, for further comparison. Although $BV(RI)_{\rm C}$ photometry 
provides a more complete coverage of the spectral energy distribution, 
\cite{c10} verified that there is essentially no difference with respect to 
using only $B_TV_T$ Tycho2 (at least for temperatures $> 5000$~K), 
the near-infrared calibration still being the dominant ingredient. 

For stars in common the difference in the IRFM (Johnson-Cousins--SAAO minus 
Tycho2--SAAO) is
$\Delta \teff=-25\pm 6$~K ($\sigma=31$~K), 
$\Delta \theta=0.3 \pm 0.1$ per cent ($\sigma=0.4$),
$\Delta \fbol=-1.0\pm0.2$ per cent ($\sigma=1.2$) per cent, 
i.e.\ the IRFM in the Johnson-Cousins--SAAO system returns slightly cooler 
temperatures, lower bolometric fluxes and larger angular diameters. 
A similar $-25$~K offset was found by \cite{c12} using Johnson-Cousins--2MASS 
versus Tycho2--2MASS photometry.

From the SAAO over 2MASS calibration in the previous Section, and the Tycho2 
versus Johnson comparison carried out here, we can conclude that changing 
the adopted filter set can systematically affect temperatures by $20-30$~K, 
in agreement with the zero-point uncertainty of $20$~K estimated by \cite{c10} 
and included in our global error estimate.

\subsection{Solar twins}\label{sect:twins}

The zero-point of the Tycho2--2MASS (and Johnson-Cousins--2MASS) scale of 
\cite{c10} was finely tuned to render the solar temperature on average for a 
set of 10 spectroscopically selected solar twins. These were identified with a 
purely differential analysis with respect to a solar reference spectrum, 
obtained with the same instrument and observing run, without assuming {\it a 
priori} any $\teff$ \citep{melendez07,melendez09:twins,ramirez09}, so that 
they truly serve as an independent test of the temperature scale.

Six of them have SAAO near-infrared photometry from this work; the 
corresponding IRFM temperatures are listed in Table~\ref{t:twins}. 
We also include the solar twin 18~Sco \citep[][]{demello97,bazot11} 
and HD138573 \citep{dfp12}. 
The Tycho2--2MASS and Tycho2--SAAO systems are clearly well calibrated with 
each other, with typical differences of only a few kelvin 
(median $7 \pm 6$~K, $\sigma=15$~K). The absolute solar 
calibration of our IRFM is also confirmed, as the average effective temperature
is very close to $T_{\rm{eff},\odot}$ for the Tycho2--2MASS and Tycho2--SAAO case
(while the Johnson--SAAO estimate is cooler by about 25~K, in agreement 
with Section \ref{sect:Johnson}).

\begin{table}
\centering
\caption{IRFM $\teff$ for solar twins. In both cases errors do not include 
the zero-point uncertainty in the effective temperature scale 
\citep[see][]{c10}.}
\label{t:twins}
\begin{tabular}{ccccc}
\hline
       &        &{\tiny Tycho2--SAAO}&{\tiny Tycho2--2MASS}&{\tiny JC--SAAO} \\
 HD    & HIP    &      $\teff$    &      $\teff$    &   $\teff$       \\
\hline
71334  & 41317  &  $5741 \pm  22$ &  $5739 \pm  27$ & $5719 \pm 32$   \\
78534  & 44935  &  $5774 \pm  29$ &  $5803 \pm  30$ & $5790 \pm 36$   \\
78660  & 44997  &  $5784 \pm  25$ &  $5791 \pm  30$ & $5726 \pm 30$   \\
115169 & 64713  &  $5822 \pm  36$ &  $5853 \pm  36$ & $5740 \pm 32$   \\
142331 & 77883  &  $5670 \pm  30$ &  $5660 \pm  35$ & $5674 \pm 33$   \\
167060 & 89650  &  $5861 \pm  29$ &  $5864 \pm  35$ & $5823 \pm 34$   \\
146233 & 79672  &  $5819 \pm  26$ &                 & $5789 \pm 38$   \\
138573 & 76114  &  $5765 \pm  23$ &  $5777 \pm  25$ &                 \\
\hline
\multicolumn{2}{c}{weighted mean} & $5778 \pm 20$ & $5782 \pm 25$ & $5750 \pm 19$ \\
\hline
\end{tabular}
\end{table}

\section{Testing the zero-point versus interferometry}
\label{sect:zeropoint}

The main purpose of this work is to compare angular diameters derived from 
the IRFM to interferometric measurements, so as to check the IRFM 
diameter (and thus temperature) scale.
While such a test is certainly fundamental, we warn that the 
calibration of the interferometric scale is not straightforward to assess:
comparisons between diameters obtained from PTI and CHARA suggest possible 
systematics at the $6 \pm 6$ per cent level \citep{vBvB09,bo12a}, corresponding
to about $150$~K in temperature; and systematics appear even when the same 
interferometer but different beam combiners are used 
\citep[e.g.,][]{bo12b,w13}. 
Although the evidence is as yet poorly quantified, it is very important 
if we aim at setting the zero point of the $\teff$ scale to better than one 
per cent (i.e.~about 2 per cent in angular diameters).
Also, the correction 
from uniform disk measurement to limb-darkened diameter plays a role. 
In particular, 3D model atmospheres provide a more realistic description 
of the centre-to-limb variation \citep{p13} and the resulting limb-darkening 
coefficients imply angular diameters smaller by $0.5-1.0$ per cent,
the exact value depending on the parameters of the star analyzed, as 
well as on the wavelength of observation
\citep[e.g.,][]{allende02,auf05,bigot06,chi10,chi12}. 
All the stars used in this paper have interferometric limb-darkened diameters 
computed using 1D models (with the exception of HD122653 discussed below in
Section~\ref{sect:giants}), and likely to be overestimated by the amount 
mentioned above.

\subsection{Comparing angular diameters}

The fundamental comparison with interferometry relies on angular diameters
(directly measured) rather than on effective temperatures, that are secondary
quantities obtained by combining the above measurements with a reconstruction 
of bolometric fluxes. 
From our sample of stars having SAAO infrared photometry we searched recent 
literature looking for interferometric measurements (first part of 
Table~\ref{tab:AD}; additional stars in the same table with unsaturared 2MASS 
photometry are discussed later).

Cherry-picking single measurements from the literature would allows us
a degree of freedom difficult to assess, and indeed we verified that with 
``appropriate'' choices on the dataset to consider, the offsets discussed 
later can essentially reduce to zero. To avoid such a bias, we assemble a 
blind sample: for stars having multiple measurements we computed their 
weighted average 
with weights $w_i=1/\sigma^2_i$, where $\sigma_i$ is the quoted uncertainty 
of each measure. Multiple measurements allow us to estimate realistic error 
bars via the weighted sample variance 
$\frac{\sum_i w_i (x_i-\mu)^2}{\sum_i w_i}$, which essentially measures the 
overdispersion of the data with respect to the simple variance of the weighted 
mean $1/\sum_i w_i$. This simple exercise suggests that --- for the sample 
here available --- realistic interferometric error bars 
for measurements below $1$~mas should be $\gtrsim 0.01$~mas.

On the contrary, the accuracy of the angular diameters determined via IRFM 
stays constant at the $1-2$ per cent level independently of the size of the 
star. This means that the best regime for testing the $\teff$ scale is by 
using stars with diameters of order $\gtrsim 1$~mas. One should not disregard, 
however, the most recent measurements with spectacular sampling of 
visibilities achieving indeed robust sub-milliarcsec results \citep{w13}. 

\begin{table*}
\centering
\caption{Angular diameters and effective temperatures measured from 
interferometry and obtained from the IRFM in different systems. The first 16 
stars are from our SAAO sample, while the remaining are in the 2MASS system 
(see discussion in the text).}\label{tab:AD}
\begin{tabular}{rcccccccc}
\hline 
            &                    &      &\multicolumn{2}{c}{Tycho2--SAAO}  &\multicolumn{2}{c}{Johnson-Cousins--SAAO}&\multicolumn{2}{c}{Tycho2--2MASS}\\ 
\hline
      HD    &      $\theta$      & Ref. &   $\teff$     &       $\theta$   &     $\teff$    &       $\theta$    &     $\teff$    &       $\theta$    \\ 
\hline 
     30652  &  $1.525 \pm 0.010$ & 1,2 & $6536 \pm  59$ & $1.511 \pm 0.021$ & $6493 \pm  62$ & $1.519 \pm 0.021$ &       $-$      &        $-$        \\
     39587  &  $1.053 \pm 0.011$ & 1,2 & $5972 \pm  40$ & $1.053 \pm 0.013$ &         $-$    &      $-$          &       $-$      &        $-$        \\
     48737  &  $1.401 \pm 0.009$ & 2   & $6553 \pm  60$ & $1.381 \pm 0.018$ &         $-$    &      $-$          &       $-$      &        $-$        \\
     56537  &  $0.835 \pm 0.013$ & 2   & $8346 \pm 136$ & $0.770 \pm 0.016$ &         $-$    &      $-$          &       $-$      &        $-$        \\
     75732  &  $0.711 \pm 0.004$ & 3   & $5295 \pm  36$ & $0.689 \pm 0.008$ &         $-$    &      $-$          &       $-$      &        $-$        \\
     97603  &  $1.324 \pm 0.021$ & 1,2 & $8115 \pm 123$ & $1.298 \pm 0.025$ &         $-$    &      $-$          &       $-$      &        $-$        \\
    102870  &  $1.433 \pm 0.006$ & 2,4 & $6146 \pm  45$ & $1.419 \pm 0.016$ & $6118 \pm  56$ & $1.423 \pm 0.017$ &       $-$      &        $-$        \\
    118098  &  $0.852 \pm 0.009$ & 2   & $8240 \pm 129$ & $0.858 \pm 0.018$ & $8154 \pm 118$ & $0.863 \pm 0.018$ &       $-$      &        $-$        \\
    131977  &  $1.177 \pm 0.030$ & 5   & $4614 \pm  40$ & $1.156 \pm 0.016$ & $4614 \pm  38$ & $1.157 \pm 0.013$ &       $-$      &        $-$        \\
    141795  &  $0.768 \pm 0.017$ & 2   & $8287 \pm 129$ & $0.729 \pm 0.014$ & $8203 \pm 117$ & $0.732 \pm 0.013$ &       $-$      &        $-$        \\
    142860  &  $1.217 \pm 0.005$ & 1,2 & $6345 \pm  56$ & $1.197 \pm 0.018$ &         $-$    &      $-$          &       $-$      &        $-$        \\
    146233  &  $0.676 \pm 0.006$ & 6   & $5819 \pm  41$ & $0.671 \pm 0.010$ & $5789 \pm  53$ & $0.674 \pm 0.010$ &       $-$      &        $-$        \\
    164259  &  $0.775 \pm 0.027$ & 2   & $6809 \pm  74$ & $0.713 \pm 0.011$ & $6777 \pm  75$ & $0.716 \pm 0.011$ &       $-$      &        $-$        \\
    173667  &  $1.000 \pm 0.006$ & 2   & $6425 \pm  58$ & $0.980 \pm 0.014$ &         $-$    &      $-$          &       $-$      &        $-$        \\
    177724  &  $0.895 \pm 0.017$ & 2   & $9152 \pm 122$ & $0.888 \pm 0.014$ &         $-$    &      $-$          &       $-$      &        $-$        \\
    182572  &  $0.845 \pm 0.025$ & 2   & $5550 \pm  36$ & $0.870 \pm 0.009$ & $5537 \pm  46$ & $0.871 \pm 0.010$ &       $-$      &        $-$        \\
\hline
       173701  &  $0.332 \pm 0.006$ & 7   &       $-$      &        $-$        &         $-$    &      $-$          & $5357 \pm  91$ & $0.324 \pm 0.013$ \\ 
       175726  &  $0.346 \pm 0.007$ & 7   &       $-$      &        $-$        &         $-$    &      $-$          & $6079 \pm 120$ & $0.346 \pm 0.016$ \\
$^{(g)}$175955  &  $0.680 \pm 0.010$ & 7   &       $-$      &        $-$        &         $-$    &      $-$          & $4766 \pm  97$ & $0.656 \pm 0.032$ \\
$^{(g)}$177151  &  $0.570 \pm 0.010$ & 7   &       $-$      &        $-$        &         $-$    &      $-$          & $5016 \pm  84$ & $0.535 \pm 0.021$ \\
       177153  &  $0.289 \pm 0.006$ & 7   &       $-$      &        $-$        &         $-$    &      $-$          & $6063 \pm 115$ & $0.279 \pm 0.012$ \\
$^{\dag}$181420  &  $0.340 \pm 0.010$ & 7   &       $-$      &        $-$        &         $-$    &      $-$          & $6637 \pm 129$ & $0.307 \pm 0.014$ \\
$^{(g)}$181827  &  $0.473 \pm 0.005$ & 7   &       $-$      &        $-$        &         $-$    &      $-$          & $4997 \pm  92$ & $0.490 \pm 0.022$ \\
       182736  &  $0.436 \pm 0.005$ & 7   &       $-$      &        $-$        &         $-$    &      $-$          & $5205 \pm  98$ & $0.443 \pm 0.020$ \\
       187637  &  $0.231 \pm 0.006$ & 7   &       $-$      &        $-$        &         $-$    &      $-$          & $6290 \pm 115$ & $0.223 \pm 0.010$ \\
$^{(g)}$189349  &  $0.420 \pm 0.006$ & 7   &       $-$      &        $-$        &         $-$    &      $-$          & $5000 \pm  89$ & $0.441 \pm 0.018$ \\
\hline 
            &                    &     &                &                   &\multicolumn{2}{c}{Johnson-Cousins--2MASS}&          &                   \\ 
    122563  &  $0.940 \pm 0.011$ & 8  &       $-$      &        $-$        & $4600 \pm  47$ & $0.941 \pm 0.019$ &       $-$      &        $-$        \\
\hline
\end{tabular}
\begin{list}{}{}
\item[] 
  Source of interferometric measurement. In case of multiple entries, weighted 
  average is taken (see text for details). $1-$ \cite{vBvB09}; 
  $2-$ \cite{bo12a}; $3-$ \cite{vB11}; $4-$ \cite{north09}; 
  $5-$ \cite{d09}; $6-$ \cite{bazot11}; 
  $7-$ \cite{huber12}, with prefix (g) for giants; $8-$ \cite{creevey12}. 
  $^{\dag}$ Interferometric measurement to be regarded with suspicion 
  due to problem with the calibrator (Daniel Huber, private communication). 
\end{list}
\end{table*}

\begin{figure*}
\begin{center}
\includegraphics[width=0.99\textwidth]{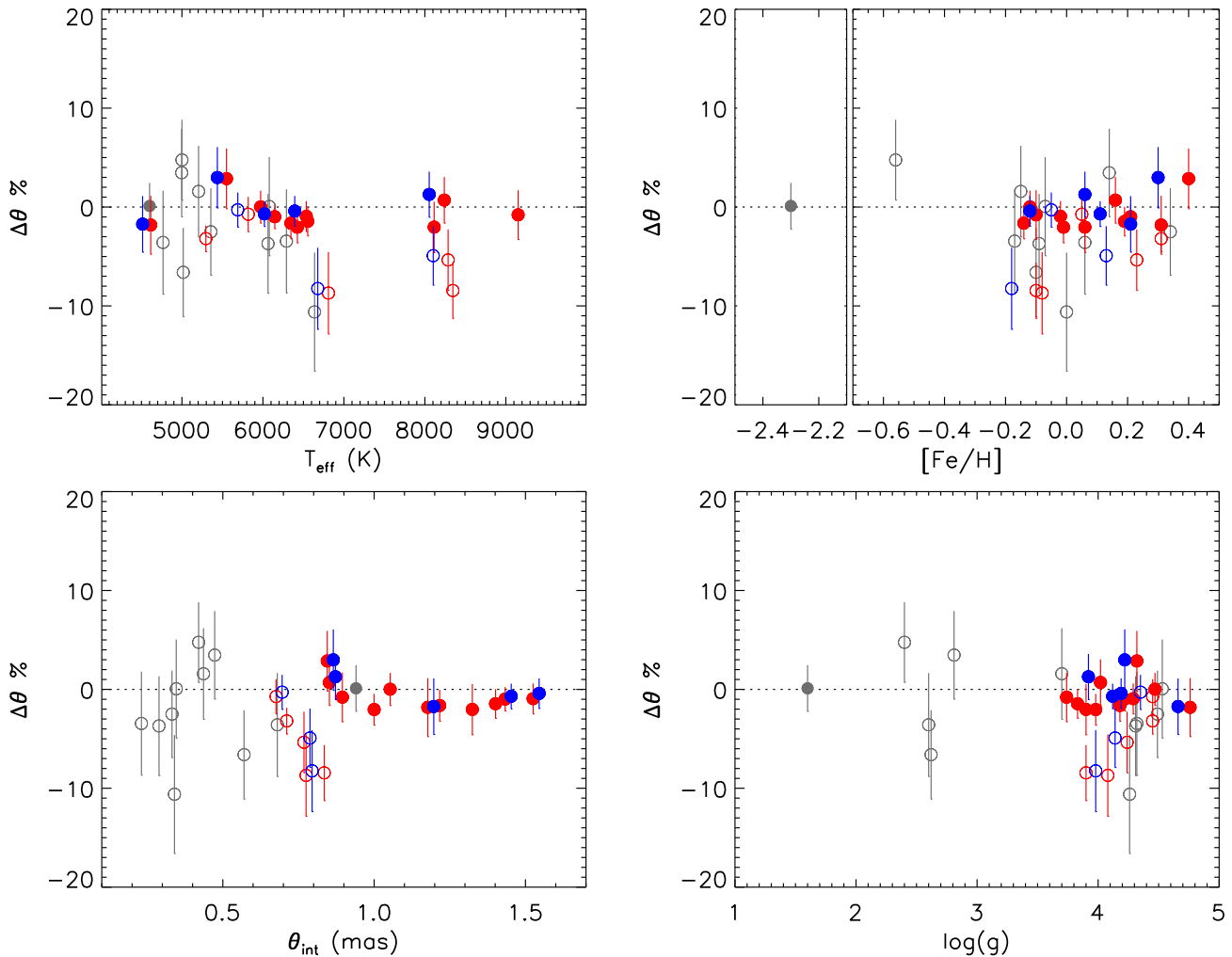}
\caption{Difference between interferometric angular diameters and 
those derived via the IRFM 
($ \Delta \theta = 1 - \theta_{\rm{int}}/ \theta_{\rm{IRFM}}$ in 
per cent). Filled symbols are for stars with interferometric diameters greater 
than $0.84$~mas (threshold above which the scatter with respect to 
interferometry is minimized, c.f.~Figure \ref{fig:d2} and discussion in the 
text).
Colours refer to the photometric systems implemented in the IRFM: 
Tycho2--SAAO in red, Johnson-Cousins--SAAO in blue (abscissa values slightly 
shifted for clarity). Gray open circles are stars from Huber et al.\,(2012) and 
the filled gray dot is HD122563.}
\label{fig:int}
\end{center}
\end{figure*}

\begin{figure}
\begin{center}
\includegraphics[width=0.5\textwidth]{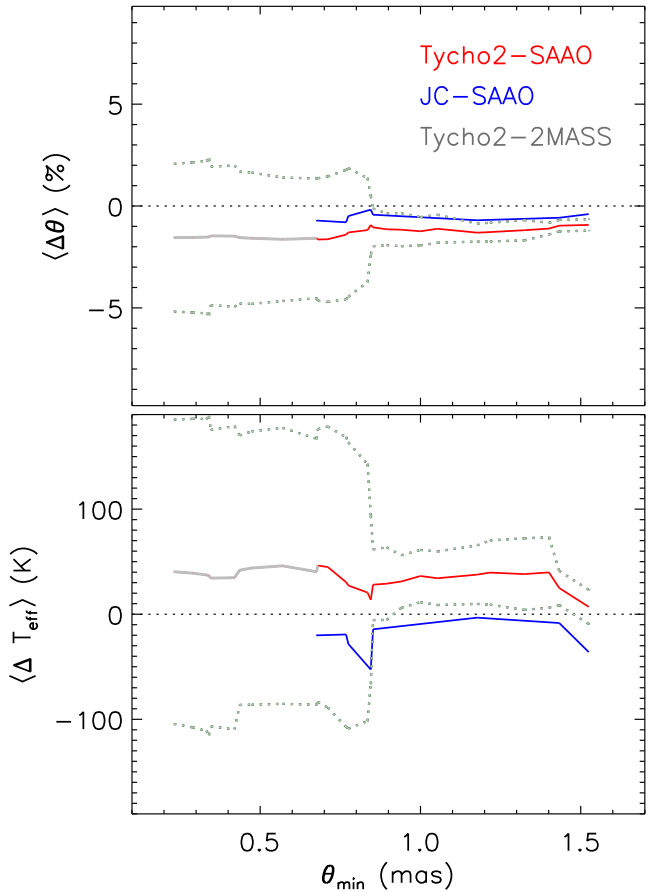}
\caption{TM-Diagram. {\it Top panel:} weighted mean difference (in per cent) 
  between angular diameters from interferometry and the IRFM as function of 
  the threshold above which interferometric diameters are considered 
  ($\theta_{\rm{min}}$). 
  Dotted lines are $1 \sigma$ error bars scatter for the Tycho2--SAAO/2MASS 
  comparison. 
  {\it Bottom panel:} same as above, but for weighted mean difference in 
  effective temperatures $\langle\Delta\teff\rangle$.} 
\label{fig:d2}
\end{center}
\end{figure}

Figure \ref{fig:int} shows the difference between IRFM and interferometric 
angular diameters, as a function of temperature, diameter, metallicity and 
gravity. No significant trend is detected and the offsets are of a pure 
zero-point nature, as appropriate for the IRFM.
From the bottom left panel, it is clear that 
the comparison is most meaningful at $\theta \gtrsim 0.8-0.9$~mas; scatter and 
uncertainties significantly increase when including data with smaller 
diameters. This is highlighted in the top panel of Figure \ref{fig:d2}, where 
the weighted mean difference (and corresponding scatter) between the IRFM and 
interferometric results is plotted as function of the threshold above which 
angular diameters are considered: $\theta_{\rm{min}}$. The $\theta_{\rm{min}}$ 
diagram (TM-Diagram) first introduced here is a handy graphic 
diagnostic to assess whether a given interferometric dataset is affected by 
systematic trends. In fact, while interferometric measurements are the more 
challenging the smaller the diameter, a photometric technique like the IRFM is 
insensitive to that. The plot also visualizes how stable the comparison is, 
when relying on just a few high quality datapoints (right-hand side of the 
plot) or a larger dataset including smaller and less precise measurements. We 
will use this test again in Section \ref{sect:GCS}.

We queried the Simbad database to flag possible troublesome stars; HD56537 and 
HD173667 
include two components (thus affecting their photometry), while HD118098 has 
a faint stellar companion.
In the latter case, the difference in $H$ band is 7 magnitudes \citep{hink10}, 
so the contribution to the flux can be readily computed to be of order 
$0.1$ per cent, i.e.~ten times smaller than observational uncertainties. 
In the case of HD56537 and HD173667, we carried out the comparison with and 
without these two stars: for the culled sample the offsets are slightly 
reduced, but the change is marginal so, for simplicity, we report only 
the results of the full sample. Similarly, our IRFM sample reflects the status 
of a rapidly evolving literature, and it does not include any of the robust 
sub-milliarcsec measurements of White et al.\,(2013, their 16\,Cyg\,B having 
good 2MASS photometry), while we learnt that the interferometric measurement of 
HD181420 should be disregarded due to problems with the calibrator (Daniel 
Huber, private communication). Both updates would improve the agreement 
between the IRFM and interferometry, but in the spirit of a blind sample we 
refrain from playing any minor star-in/star-out game to the sample.   
Also, we find very similar offsets when using other statistical estimators
(simple mean or median, rather than weighted mean).

All in all, in the regime most relevant for the comparison 
($\theta \gtrsim 0.8-0.9$~mas), we find offsets in diameters of order 
$-1.0$ per cent for the Tycho2--SAAO IRFM stars, and $-0.5$ per cent for 
the Johnson-Cousins--SAAO IRFM stars. This is consistent with the average 
offset of $-0.62 \pm 1.7$ per cent found in \cite{c10} applying $\teff$ and 
bolometric flux calibrations to an almost 
entirely different set of stars.
\footnote{Only three of the stars in the present sample 
overlap with those used in \cite{c10} (c.f.~their Table 3): HD75732, HD102870 
and HD131977, all in excellent agreement within errors.} 
Thus the comparisons performed in that work were sound, albeit indirect.

The above offsets in angular diameters translate into $\teff$ hotter by 
$\sim 0.5$ and $0.25$ per cent, i.e.\ by $\sim +30$~K and $\sim +15$~K at the 
solar value. We conclude that, depending on the exact filter set used, the 
IRFM scale agrees with interferometry on average within about 20~K, as 
originally claimed.
Such offset is comparable to the systematic change expected when adopting 
3D limb--darkening corrections in interferometric measurements 
\citep[e.g.,][]{allende02,chi10}, possibly bringing the 
two scales in even closer agreement. Considering the systematics involved in 
interferometric measurements (see also Section \ref{sect:GCS}), one may 
actually argue that the reliability of 
the latest absolute fluxes \citep[at the per cent level,][]{bohlin07} and the 
solar twin calibration, rival interferometry in setting the 
temperature scale. The offsets we find are in fact at a level where 
also interferometric measurements are plagued by systematics. 

\subsection{Comparing temperatures}

The comparison in terms of directly measured angular diameters is the most 
robust; comparison to interferometric temperatures is less straightforward, 
for these involve an additional reconstruction of the bolometric flux 
As the flux derivation is independent in the two approaches, the comparison 
between temperatures may introduce additional noise and systematics. In 
particular, interferometric papers often allow reddening as a free parameter 
in the fit of the spectral energy distribution 
\citep[e.g.,][]{vBvB09,bo12a}; while for stars as nearby as those in our 
sample an assumption of negligible reddening is more appropriate.

Nonetheless, as it is customary for the various temperature scales in the 
literature to be compared to interferometric $\teff$'s, in the bottom panel of 
Figure \ref{fig:d2} we also perform this check for the stars in Table 
\ref{tab:AD} where literature reports interferometric temperatures along with 
diameters (the majority) . 
Reassuringly, the resulting temperature offsets are consistent with the 
more direct comparison of diameters in the previous section: the independent 
reconstruction of $\fbol$ in the two methods does not result in any 
significant systematic differences.

\subsection{Giants, and the benchmark metal-poor case: HD122563}
\label{sect:giants}

The increasing capabilities of interferometers are pushing the limit to which 
angular diameters can now be measured, in particular using optical beam 
combiners. Recently, \cite{huber12} measured angular diameters $<0.7$~mas
for a number of stars with unsaturated 2MASS photometry; these stars are
listed in the second part of Table~\ref{tab:AD} and included in 
Figure \ref{fig:int} (gray open circles) and \ref{fig:d2} (gray lines).
Their main conclusion 
is that angular diameters from interferometry and the IRFM agree within 
$-2 \pm 2$ per cent ($\sigma = 5$ per cent); or $-1.4 \pm 1.5$ per cent in 
weighted average. This agrees with what we found in the previous Section 
for our SAAO stars with larger and more accurate diameters.
More interestingly, four stars in \cite{huber12} are giants; 
their diameters agree very well with our estimates:
$0.0 \pm 2.3$ per cent ($\sigma = 5.5$ per cent) in weighted average. 

Because of its relevance as a standard metal-poor star, we include in the 
discussion also HD122563 for which the angular diameter has been recently 
measured, and corrected using 3D limb-darkening 
coefficients \citep{creevey12}. As there is no SAAO $JHK$ photometry for 
this star and its 2MASS data are saturated, for this one star we resort 
to filter transformations from Johnson $JHK$ to 2MASS using the updated 
\cite{carpenter01} transformations; Johnson--Cousins photometry is used in the 
optical. The star is a giant located at 
$\sim 240$~pc \citep{vanLeeuwen07} and reddening must be considered. We adopt
$A_v =0.01$ from \cite{creevey12}, which corresponds to $E(B-V)=0.003$, in 
very close agreement with estimations from interstellar NaD lines in 
high-resolution spectra of this star (Maria Bergemann, private communication). 
Thus reddening has a minor contribution to the global error budget, $\teff$ 
being only $8$~K cooler should attenuation be zero.
This metal-poor star is displayed in Figure \ref{fig:int} as a filled gray dot:
the agreement between IRFM and interferometric results is near perfect.
Together with a former test using HST absolute spectrophotometry \citep{c10},
this result confirms the robustness of the IRFM also at low metallicities. 
Setting the $\teff$ scale in the metal-poor regime is crucial, for 
an uncertainty of $100$~K can rival with NLTE effects on determinations 
of iron abundance \citep{Ruchti2013}, and affect the lithium level on 
the Spite plateau, with cosmological implications 
\citep[e.g.,][]{melendez09:lithium,sbo10,nor12}.

\section{The interferometric scale: a cautionary tale}
\label{sect:GCS}

One of the original motivations behind this work was to test the temperature 
scale of the Geneva-Copenhagen Survey. The latest official
rendition \citep[GCSIII,][]{holmberg09}\footnote{For convenience in the 
following we refer to GCSIII, although stellar parameters 
were derived in GCSII, see discussion in section 2.1.1 of \cite{c11}.} 
and the independent analysis of 
the same sample based on the IRFM scale differ by about 80~K 
\citep[][hereinafter C11, being hotter]{c11}; yet, both authors claim 
agreement with interferometry within the errors.
It was pointed out that considering a dozen stars in common between the 
Geneva-Copenhagen catalogue and the recent CHARA dataset of \citet{bo12a} with 
$\theta \geq 1$~mas, GCSIII temperatures are on average in excellent agreement 
with the interferometric 
scale, while C11 temperatures are too hot by about 70~K --- at odds with our 
results so far. Even more confusingly, \cite{bo12a} reported good agreement 
with the $(B-V)$ colour--metallicity--$\teff$ relation of \cite{c10}, but 
found GCSIII temperatures to be too {\it hot} by $100$~K or more.

Spurred by this, we applied the TM-Diagram to the dataset of \cite{bo12a}, 
which has 25 stars in common with either 
GCSIII or C11\footnote{From an initial sample of 33 stars in common we exclude 
7 objects 
for which photometry includes more than one component, as marked by the 
corresponding label (usually ``AB''). As recommended in C11, users are warned 
against their use, although we verified that the overall conclusions still 
hold, should these stars be kept.}.
Note that all stars discussed in the remaining of this Section have 
effective temperatures derived from the Str\"omgren $(b-y)$ index, if not 
otherwise specified. This is the 
colour used in GCSIII for all targets, and in C11 for stars with unreliable 
2MASS photometry to apply the IRFM (indeed the case for all nearby 
interferometric targets). 

\begin{figure}
\begin{center}
\includegraphics[width=0.5\textwidth]{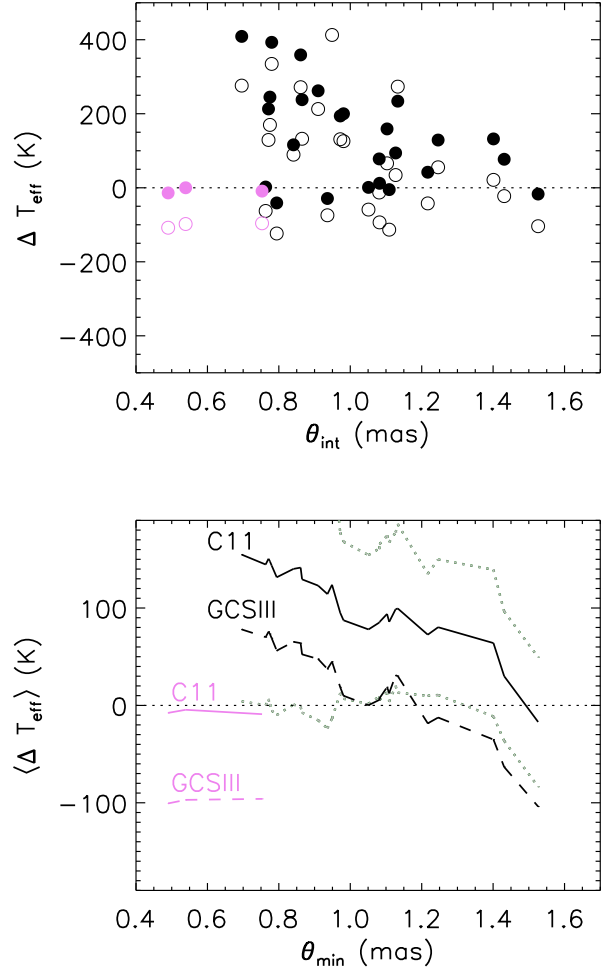}
\caption{{\it Top panel:} effective temperatures of Casagrande et al.\,(2011, 
filled circles) and Holmberg et al.\,(2009, open circles) with respect to 
interferometric measurements of Boyajian et al.\,(2012a, black) and White 
et al.\,(2013, purple). {\it Bottom panel:} weighted mean difference for stars 
in the top panel as function of the threshold above which interferometric 
angular diameters are considered (TM-Diagram). One $\sigma$ errors with 
respect to C11 are included for comparison (gray dotted lines).}
\label{fig:d1}
\end{center}
\end{figure}

In Figure \ref{fig:d1} (top panel) we show the effective temperature 
difference between the two Str\"omgen scales and the 
\cite{bo12a} dataset, plotted as function of interferometric angular 
diameters. 
A clear trend appears, which is further highlighted in the bottom panel with 
the TM-Diagram, where the average difference with respect to the above 
interferometric temperatures $\langle \Delta \teff \rangle$ is shown for 
various thresholds in angular diameters. 
While the constant difference between GCSIII and C11 reflects the zero-point 
difference of these two photometric scales (and their similar internal 
consistency), the trend with respect to the interferometric measurements of 
\cite{bo12a} suggests the presence of systematic effects in the latter sample. 
We also verified that our conclusions still hold should this plot be done with 
$\teff$ derived from colour indices other than Str\"omgren (the value of the 
offset using other colours might vary by a few tens of K --compatible with 
the intrinsic difference when using different indices on a rather 
limited number of stars-- but the trend remains).

\begin{figure*}
\begin{center}
\includegraphics[width=0.99\textwidth]{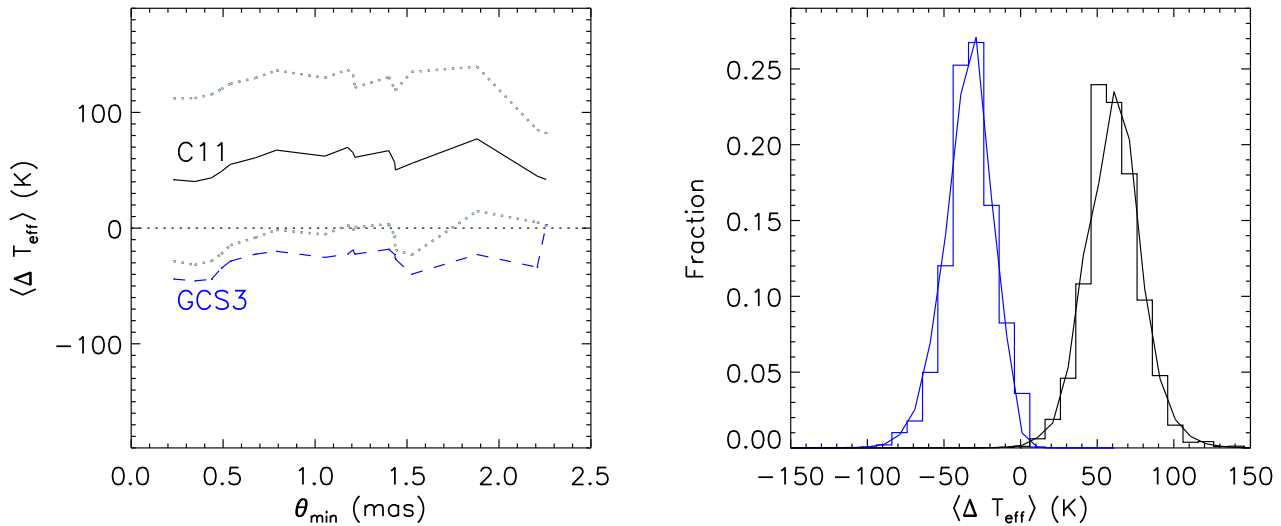}
\caption{{\it Left panel:} TM-Diagram for effective temperatures in GCSIII and 
C11 from Table 3 plus interferometric measurements other than Boyajian et 
al.\,(2012a). 
Additional interferometric sources used are: Mozurkewich et al.\,(2003), van 
Belle et al.\,(2007), North et al.\,(2007), Kervella et al.\,(2008), van 
Belle \& von Braun (2009), Demory et al.\,(2009), Huber et al.\,(2012), 
White et al.\,(2013). Only unreddened stars and with interferometric errors 
smaller than 5 per cent are used. Multiple measurements of same stars are 
averaged together as in Section \ref{sect:zeropoint}. {\it Right panel:} 
median (histogram) and mean (continuous line) distribution of the weighted 
effective temperature difference for $2\times10^{6}$ random realizations of 
the plot in the left panel.}
\label{fig:monte}
\end{center}
\end{figure*}
\nocite{north07}
\nocite{km08}
\nocite{vb07}
\nocite{moz03}

This systematic trend hampers robust conclusions on the $\teff$ zero-points 
from this dataset: the offset between GCSIII and \cite{bo12a} is null if 
restricting to $\theta_{\rm{min}} \geq 1$~mas, while when all
common stars are considered, GCSIII is about 100~K hotter. This was indeed
the claim in \cite{bo12a}, and it would consequently imply that the C11 scale 
is too hot by about 200~K, which would be very surprising in view of the 
results in Section~\ref{sect:zeropoint} for our SAAO sample, and the solar 
twins test.
We note further that the results of \cite{bo12a} for the famous solar twin 
18~Sco (for which they obtain $\theta=0.780 \pm 0.017$~mas, implying 
$R=1.166 \pm 0.026 \,R_\odot$ and $\teff =5433 \pm 69$~K) fits the systematic 
trend highlighted in Figure~\ref{fig:d1}; while \citet{bazot11} with PAVO 
measure $\theta=0.676 \pm 0.0062$~mas and confirm the strictly solar radius 
($R=1.010 \pm 0.009 \,R_\odot$), mass and $\teff$ of this star.

It is clearly hard to objectively set a threshold above which the comparison 
can be considered meaningful, and other interferometric 
measurements are required to gauge this problem. \cite{w13} highlights 
underestimated errors and systematic offsets in the sample of \cite{bo12a}.
Both studies are carried out with CHARA, but the latter uses the PAVO 
optical beam combiner instead of Classic. PAVO allows to probe the visibility 
curve at higher spatial frequencies, which are needed to derive robust angular 
diameters
\citep[see fig.~3 in][]{w13}. For one of the targets in \cite{w13}, 
\cite{Ligi2012} independently confirms a diameter significantly smaller, and 
thus a hotter $\teff$, than \cite{bo12a}.
Considering the stars of \cite{w13} in Figure \ref{fig:d1} shows no trends
with diameter, as one would expect, and yields good agreement of C11 with 
this particular interferometric set.  

In retrospective, we stress that no trend is present in Figure~\ref{fig:d2}, 
based on a compilation from literature \citep[including][refraining from its 
exclusion in the spirit of a blind sample]{bo12a}, but averaging over multiple 
measurements of the same star when available. We also remark that in 
Figure~\ref{fig:d2} the scatter at all $\theta_{\rm{min}}$ is also much lower 
than in Figure \ref{fig:d1}, and excluding the dataset of \cite{bo12a} from 
Table \ref{tab:AD} does not change significantly the conclusions of Section 
\ref{sect:zeropoint}.

The suspicion that there are systematic trends in the \cite{bo12a} dataset is 
further highlighted once the same comparison is performed with respect to 
other measurements in literature (see caption of Figure \ref{fig:monte}). 
Over a wide range of angular diameters, this 
comparison shows a rather constant offset in effective temperature (left panel 
in Figure \ref{fig:monte}). It also confirms the known offset of 
about $80$~K between GCSIII and C11, and it shows how the two scales are 
compatible with interferometry within $\pm 50$~K or better, on the cool and 
hot side respectively. This is not entirely surprising, now: the 
effective temperatures used for this comparison are in fact 
all derived from the $(b-y)$ colour. 
This Str\"omgren index was calibrated against 
$\teff$ derived in a more direct way (the ``parent scales''):
the IRFM in C11 and the \cite{diBene98} surface brightness relation in GCSIII. 
We verified from over 70 common stars, that the parent scales differ by 
about $40$~K. Consistently with the offsets found in 
Section~\ref{sect:zeropoint} for the IRFM, the \cite{diBene98} scale is in 
good agreement with interferometry, but on the coolish side ($-9 \pm 22$, 
$\sigma=108$~K, for 25 of his stars with modern interferometric data).
Thus, the same must be true for effective temperatures derived from their 
colour calibrations, but with important caveats. 
We verified that {\it on average} the effective temperatures derived from the 
$(b-y)$ index are both excellent renditions of the corresponding parent scales,
although when considering a limited number of stars, 
zero-point differences of a few tens of Kelvin are possible 
\citep[cf.\,also][]{munoz13}. 
These differences stem both from the fact that the colour relations are not 
always a perfect rendition of the calibrating sample over the full parameter 
space, as well as from the photometric errors associated to measurements in 
each colour index.
In particular, the sensitivity of $(b-y)$ to metallicity as well as 
its steep correlation with $\teff$ make this colour index less than optimal 
for discriminating the zero-point of various effective temperature scales: a 
change of only $0.010-0.015$ mag in $(b-y)$ corresponds to a shift of $100$~K, 
a change which can be as small as $0.007$ mag if the joint effect of 
metallicity is included. 
Thus, it is not surprising that in Figure \ref{fig:monte} the $(b-y)$ effective 
temperatures of both GCSIII and C11 have a somewhat different offset than  
expected from the parent scales (di Benedetto or the direct IRFM comparison of 
Section \ref{sect:zeropoint}).

Our tests show that the offsets derived for colour--calibrated
temperature scales are quite sensitive to the specific subsample of
interferometric stars considered for comparison, and we illustrate this
with a Monte Carlo simulation in the right panel of Figure \ref{fig:monte}.
To this 
purpose, we have run $2\times10^{6}$ different realization of the sample used 
in the left panel. For each realization we took a sub-sample random in 
number of entries. The plot 
shows the distribution of the median and mean weighted average difference of 
GCSIII and C11 with respect to these random interferometric sub-samples. 
Depending on the sub-sample considered, somewhat different zero-point values 
are inferred from such a comparison. This, together with the lower accuracy 
introduced when working with $\teff$ 
derived from colour relations explain why both GCSIII and C11 are still 
compatible with interferometry, while differing between them by as much as 
$80$~K. 

Since the two scales tie in this comparison, we shall comment on 
the scope of the adopted effective temperatures in GCSIII and C11, i.e. to 
determine stellar ages. To this purpose, in GCSIII the cooler effective 
temperatures were compensated by shifting the temperatures of the reference 
stellar isochrones, effectively erasing the difference between the two 
empirical scale. The hotter $\teff$ scale adopted in C11 also inspired a 
revision of the metallicity calibration, selecting spectroscopic measurements 
with temperatures consistent with the IRFM. We notice that fifteen stars from 
the spectroscopic sources selected to calibrate metallicities in C11 can be 
now directly compared to {\it interferometric} temperatures from Figure 
\ref{fig:monte}, showing good agreement for the adopted spectroscopic scale 
$\Delta\teff = -2 \pm 25$~K ($\sigma=92$~K).

\section{Constraining $\teff$ with parallaxes}\label{sect:parallaxes}

Recently, asteroseismology has provided an alternative method to determine
stellar masses and radii via scaling relations \citep[e.g.,][]
{hekker09,stello09,vsa11}; the latter depend only loosely on the adopted 
$\teff$, and combination of asteroseismic radii with reliable angular 
diameter measurements can in principle yield {\it Hipparcos}--quality distances
\citep{vsa12}.

While recent literature has focused on testing the scaling relations, there
is by now a body of evidence suggesting that the latter are robust for 
main-sequence and subgiant stars \citep[e.g.,][and references therein]{cm13}, 
so that we dare turn the argument around and use distances (parallaxes) to test 
the angular diameter and $\teff$ scale. 
This is still limited to a handful of stars, but here we propose 
the method with a view to application to future Gaia targets.

\begin{figure*}
\begin{center}
\includegraphics[width=0.99\textwidth]{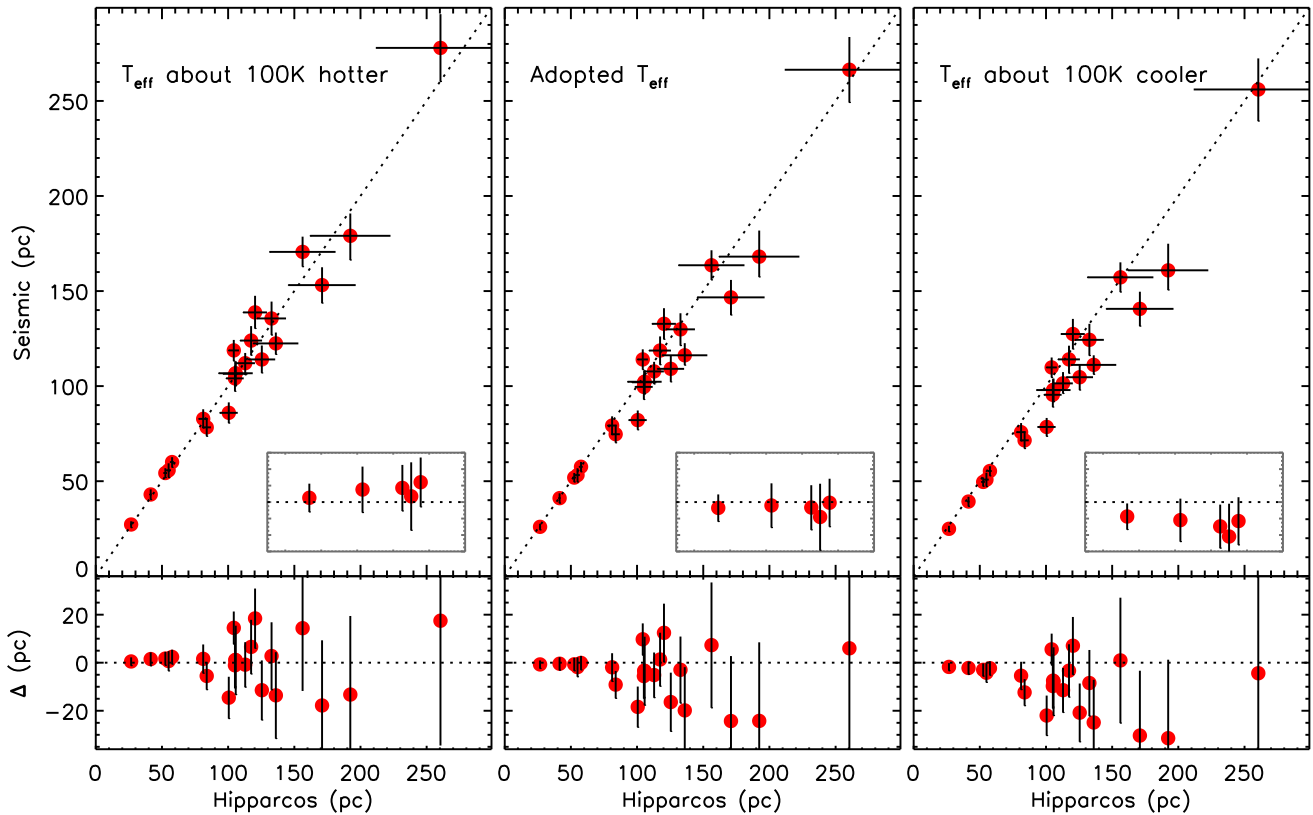}
\caption{Seismic distances of main-sequence and subgiant stars relying on our 
$\teff$ and angular diameter scale compared to {\it Hipparcos} distances. The 
effect of changing the infrared absolute calibration and thus the $\teff$ 
scale by roughly $\pm$100~K is shown on the left and right panels. 
Beyond $60$~pc, uncertainties largely increase due to reddening. Insets in the 
top panels zoom onto stars closer than $70$~pc where reddening is not an 
issue.}
\label{fig:hip}
\end{center}
\end{figure*}

In Figure \ref{fig:hip} we compare {\it Hipparcos} distances to seismic ones, 
obtained by combining our (Tycho--2MASS) IRFM temperatures and angular 
diameters with asteroseismic radii. The latter are computed as described in 
\cite{vsa12}, the only improvement with respect to the values published there 
being the use of updated asteroseismic parameters \citep{chap13} --- although 
the effect is negligible on the scale of Figure \ref{fig:hip} and we verified 
the same conclusions hold, should the previous set of frequencies be used.

As we have already extensively discussed, changing the near-infrared absolute 
calibration adopted in the IRFM returns 
different $\teff$, bolometric fluxes and angular diameters. Here we explore 
the effect of changing the 2MASS absolute calibration by $\pm 5$ per cent, 
which implies a change of about $\mp 100$~K in effective temperatures and 
$\pm 3$ per cent in angular diameters. 
It can be immediately appreciated that both a hotter or a cooler effective 
temperature scale return worse agreement in comparison with {\it Hipparcos}. 
What can be robustly concluded is that the method implemented in \cite{vsa12} 
is the one providing better distances, an important validation for using this 
technique in studies of Galactic structure.

\section{Conclusions and future perspectives}\label{sect:Conclusions}

In this paper we performed a test of the IRFM temperature scale
by \citet{c10} versus the fundamental interferometric scale. Direct comparison
between the two scales has been hindered by the lack of accurate, 
homogeneous near-infrared photometry for the nearby stars probed by 
interferometry.
In particular, 2MASS photometry which is the basis for modern IRFM 
implementations, is saturated for these nearby stars.

The purpose of this work was to fill this gap. We present dedicated SAAO 
$JHK$ photometry for 55 stars, that allows us to directly implement 
the IRFM procedure on 16 nearby stars with measured angular diameters.
The remainder of the sample has both SAAO and excellent 2MASS photometry,
and by acting only on the absolute calibration is used to secure that the 
present SAAO--based scale matches the 2MASS--based one of \citet{c10}. 
Notice that to achieve the highest accuracy possible, no transformation 
between the two systems has been performed; rather the IRFM has been 
implemented for each system using the most reliable absolute calibration, 
zero-point and filter transmission curves. When this is done, the specific 
choice of filter set (optical: Johnson-Cousins vs.\
Tycho2; or near-infrared: SAAO vs.\ 2MASS) affects the IRFM temperatures by 
about $20-30$~K.

Comparison to interferometric data with the best direct measurements 
($\theta \gtrsim 0.8-0.9$~mas) reveals offsets between $0.5$ and $1.0$ per cent 
in angular diameter (the IRFM scale having smaller diameters and thus 
hotter $\teff$, by $15-30~$K), depending on the exact filter system in use.
These offsets are close to those indirectly derived in \cite{c10} using colour 
calibrations, thus confirming the results of that analysis.
Considering that essentially all interferometric diameters considered here have 
limb--darkening corrections computed from 1D model atmospheres, adoption
of more realistic 3D corrections is deemed to largely remove the above
mentioned difference, bringing the direct and IRFM scale into even closer
agreement. Indeed, in a future work we intend to revise interferometric 
measurements using limb-darkening corrections computed from the 3D 
hydrodynamical stellar atmosphere models with the Stagger code by 
\cite{magic13a}.

The IRFM temperature scale in this work is confirmed to be well calibrated 
on solar twins, as was the original scale of \citet{c10}. Considering the 
systematics involved, the solar twin calibration stands as a competitive 
and model independent technique which rivals interferometric data
in setting the absolute scale, and being affected by virtually no systematics. 
We also find excellent agreement with the recent interferometric diameters of 
four giant stars in \citet{huber12} and with that of HD122563 at $\feh=-2.3$
\citep{creevey12}, thus providing a positive test of the IRFM temperatures for 
giants down to the metal-poor regime.

The IRFM temperature scale by \citet{c10} inspired an independent 
reanalysis of stellar parameters in the Geneva--Copenhagen Survey. Its 
temperatures are about 80~K hotter than the original GCSIII temperatures, 
inducing various effects in the interpretation of the data \citep{c11}. 
Discriminating between the two scales was one of the motivations behind this 
work. We proved that {\it direct} application of the IRFM yields temperatures 
in very good agreement with interferometry (within 15 to 30~K) when the 
latest, more accurate interferometric data are used. Such a clear cut 
conclusion {\it on the zero-point} of the $\teff$ scale is hard to reach when 
testing only colour calibrations, necessary e.g. for large photometric surveys 
like the Geneva--Copenhagen Survey. With the additional scatter and 
metallicity dependence inevitably introduced by colour--metallicity--$\teff$ 
relations, we find that the two Str\"omgren scales in GCSIII and C11 are 
respectively on the cool and hot side of the interferometric scale --- which
itself can somewhat change with the adopted compilation of data. It is however 
encouraging that the agreement is within $\pm 50$~K or better, that is in fact 
about the 1\% accuracy we are striving for.

Nevertheless, colour-$\teff$ relations have high internal consistency, 
independently on the angular size of stars. It is shown throughout the paper 
that the comparison between photometric and interferometric effective 
temperatures as function of angular diameters is in fact able to reveal 
trends in interferometric measurements at the smallest diameters.
With a critical assessment of angular diameters available in literature, we
thus conclude that 
currently for stars with $\theta \lesssim 1$~mas particular caution must be 
used, as systematics are seen to plague some specific interferometric datasets.
This must be kept in mind e.g. when discussing discrepancies between models 
and observations of barely resolved stars. 

In particular, interferometry alone is often not enough to perform 
conclusive tests, other constraints being mandatory. This is crucial for 
e.g.\,for ensuring the correct calibration of current and future large surveys, 
be they photometric or spectroscopic. In this spirit, we have 
already used more than one method to test effective temperatures in our 
past investigations (solar twins, line-depth-ratio, absolute 
spectrophotometry).
Here we present a new one to gauge the temperature and diameter scale, by 
combining astrometric distances and asteroseismic radii to the IRFM.
As of now, the method
is tentative, yet it already suggests that significantly 
different temperatures (100~K cooler or hotter than the present IRFM scale) 
are disfavoured; and it holds promise in view of the upcoming Gaia distances 
and all-sky asteroseismic missions (e.g., K2 \citealt{whitechap}; TESS 
\citealt{ricker10}).
For instance, since the adopted $\teff$ and diameter scales are now tested on 
nearby, reddening free stars, and astrometric distances and seismic radii are 
reddening independent, a possible application will be to 
derive the values of reddening upon which each IRFM angular diameter 
agrees with that estimated from its seismic radius and astrometric 
distance, thus building 3D extinction maps of the Galaxy on a 
star-by-star basis.

\section*{Acknowledgments}

L.P. and J.D. are supported by the Academy of Finland (grant nr.~208792). 
Funding for the Stellar Astrophysics Centre is provided by The Danish National 
Research Foundation (Grant agreement no.: DNRF106). The research is supported 
by the ASTERISK project (ASTERoseismic Investigations with SONG and Kepler) 
funded by the European Research Council (Grant agreement no.: 267864). 
This work has been supported by an Australian Research Council 
Laureate Fellowship to MA (grant FL110100012).
This publication makes use of the VizieR catalogue access tool and of 
the Simbad database operated by CDS, Strasbourg, France; and of data products 
from the Two Micron All Sky Survey, which is a joint project of the University 
of Massachusetts and the Infrared Processing and Analysis Center/California 
Institute of Technology, funded by the National Aeronautics and Space 
Administration and the National Science Foundation.


\begin{thebibliography}{113}
\providecommand{\natexlab}[1]{#1}

\bibitem[{{Allende~Prieto} et~al.(2002){Allende~Prieto}, {Asplund},
  {Garc\'{\i}a L\'opez} \& {Lambert}}]{allende02}
{Allende~Prieto} C., {Asplund} M., {Garc\'{\i}a L\'opez} R.~J., {Lambert}
  D.~L., 2002, \apj, 567, 544

\bibitem[{{Alonso} et~al.(1996{\natexlab{a}}){Alonso}, {Arribas} \&
  {Martinez-Roger}}]{alonso96:irfm}
{Alonso} A., {Arribas} S., {Martinez-Roger} C., 1996{\natexlab{a}}, \aaps, 117,
  227

\bibitem[{{Alonso} et~al.(1996{\natexlab{b}}){Alonso}, {Arribas} \&
  {Martinez-Roger}}]{alonso96:teff_scale}
{Alonso} A., {Arribas} S., {Martinez-Roger} C., 1996{\natexlab{b}}, \aap, 313,
  873

\bibitem[{{Andersen}(1991)}]{and91}
{Andersen} J., 1991, \aapr, 3, 91

\bibitem[{{Asplund} \& {Garc{\'{\i}}a P{\'e}rez}(2001)}]{asplund01}
{Asplund} M., {Garc{\'{\i}}a P{\'e}rez} A.~E., 2001, \aap, 372, 601

\bibitem[{{Asplund} et~al.(2009){Asplund}, {Grevesse}, {Sauval} \&
  {Scott}}]{asplund09}
{Asplund} M., {Grevesse} N., {Sauval} A.~J., {Scott} P., 2009, \araa, 47, 481

\bibitem[{{Aufdenberg} et~al.(2005){Aufdenberg}, {Ludwig} \&
  {Kervella}}]{auf05}
{Aufdenberg} J.~P., {Ludwig} H.~G., {Kervella} P., 2005, \apj, 633, 424

\bibitem[{{Bazot} et~al.(2011)}]{bazot11}
{Bazot} M. et~al., 2011, \aap, 526, L4

\bibitem[{{Bell} \& {Gustafsson}(1989)}]{bell89}
{Bell} R.~A., {Gustafsson} B., 1989, \mnras, 236, 653

\bibitem[{{Bessell}(1990)}]{b90}
{Bessell} M.~S., 1990, \aaps, 83, 357

\bibitem[{{Bigot} et~al.(2006){Bigot}, {Kervella}, {Th{\'e}venin} \&
  {S{\'e}gransan}}]{bigot06}
{Bigot} L., {Kervella} P., {Th{\'e}venin} F., {S{\'e}gransan} D., 2006, \aap,
  446, 635

\bibitem[{{Blackwell} \& {Lynas-Gray}(1994)}]{bl94}
{Blackwell} D.~E., {Lynas-Gray} A.~E., 1994, \aap, 282, 899

\bibitem[{{Blackwell} \& {Shallis}(1977)}]{blackwell77}
{Blackwell} D.~E., {Shallis} M.~J., 1977, \mnras, 180, 177

\bibitem[{{Blackwell} et~al.(1979){Blackwell}, {Shallis} \&
  {Selby}}]{blackwell79}
{Blackwell} D.~E., {Shallis} M.~J., {Selby} M.~J., 1979, \mnras, 188, 847

\bibitem[{{Blackwell} et~al.(1980){Blackwell}, {Petford} \&
  {Shallis}}]{blackwell80}
{Blackwell} D.~E., {Petford} A.~D., {Shallis} M.~J., 1980, \aap, 82, 249

\bibitem[{{Blackwell} et~al.(1990){Blackwell}, {Petford}, {Arribas}, {Haddock}
  \& {Selby}}]{bp90}
{Blackwell} D.~E., {Petford} A.~D., {Arribas} S., {Haddock} D.~J., {Selby}
  M.~J., 1990, \aap, 232, 396

\bibitem[{{Blackwell} et~al.(1991){Blackwell}, {Lynas-Gray} \&
  {Petford}}]{blp91}
{Blackwell} D.~E., {Lynas-Gray} A.~E., {Petford} A.~D., 1991, \aap, 245, 567

\bibitem[{{Bohlin}(2007)}]{bohlin07}
{Bohlin} R.~C., 2007, in C.~{Sterken}, ed., The Future of Photometric,
  Spectrophotometric and Polarimetric Standardization. Astronomical Society of
  the Pacific Conference Series, Vol. 364, pp. 315--+

\bibitem[{{Boyajian} et~al.(2012{\natexlab{a}})}]{bo12a}
{Boyajian} T.~S. et~al., 2012{\natexlab{a}}, \apj, 746, 101

\bibitem[{{Boyajian} et~al.(2012{\natexlab{b}})}]{bo12b}
{Boyajian} T.~S. et~al., 2012{\natexlab{b}}, \apj, 757, 112

\bibitem[{{Brown} \& {Twiss}(1958)}]{bt58}
{Brown} R.~H., {Twiss} R.~Q., 1958, Royal Society of London Proceedings Series
  A, 248, 222

\bibitem[{{Carpenter}(2001)}]{carpenter01}
{Carpenter} J.~M., 2001, \aj, 121, 2851

\bibitem[{{Carter}(1990)}]{c90}
{Carter} B.~S., 1990, \mnras, 242, 1

\bibitem[{{Carter} \& {Meadows}(1995)}]{cm95}
{Carter} B.~S., {Meadows} V.~S., 1995, \mnras, 276, 734

\bibitem[{{Casagrande}(2008)}]{c08:uppsala}
{Casagrande} L., 2008, Physica Scripta Volume T, 133, 014020

\bibitem[{{Casagrande}(2009)}]{c09}
{Casagrande} L., 2009, Memorie della Societa Astronomica Italiana, 80, 727

\bibitem[{{Casagrande} et~al.(2006){Casagrande}, {Portinari} \& {Flynn}}]{c06}
{Casagrande} L., {Portinari} L., {Flynn} C., 2006, \mnras, 373, 13

\bibitem[{{Casagrande} et~al.(2008){Casagrande}, {Flynn} \& {Bessell}}]{c08}
{Casagrande} L., {Flynn} C., {Bessell} M., 2008, \mnras, 389, 585

\bibitem[{{Casagrande} et~al.(2010){Casagrande}, {Ram{\'{\i}}rez},
  {Mel{\'e}ndez}, {Bessell} \& {Asplund}}]{c10}
{Casagrande} L., {Ram{\'{\i}}rez} I., {Mel{\'e}ndez} J., {Bessell} M.,
  {Asplund} M., 2010, \aap, 512, A54+

\bibitem[{{Casagrande} et~al.(2012){Casagrande}, {Ram{\'{\i}}rez},
  {Mel{\'e}ndez} \& {Asplund}}]{c12}
{Casagrande} L., {Ram{\'{\i}}rez} I., {Mel{\'e}ndez} J., {Asplund} M., 2012,
  \apj, 761, 16

\bibitem[{{Casagrande} et~al.(2011)}]{c11}
{Casagrande} L., {Sch{\"o}nrich} R., {Asplund} M., {Cassisi} S.,
  {Ram{\'{\i}}rez} I., {Mel{\'e}ndez} J., {Bensby} T., {Feltzing} S., 2011,
  \aap, 530, A138

\bibitem[{{Celis}(1986)}]{c86}
{Celis} S.~L., 1986, \apjs, 60, 879

\bibitem[{{Chaplin} \& {Miglio}(2013)}]{cm13}
{Chaplin} W.~J., {Miglio} A., 2013, \araa, 51, 353

\bibitem[{{Chaplin} et~al.(2013)}]{whitechap}
{Chaplin} W.~J. et~al., 2013, astro-ph/1309.0702

\bibitem[{{Chaplin} et~al.(2014)}]{chap13}
{Chaplin} W.~J. et~al., 2014, \apjs, 210, 1

\bibitem[{{Chiavassa} et~al.(2010){Chiavassa}, {Collet}, {Casagrande} \&
  {Asplund}}]{chi10}
{Chiavassa} A., {Collet} R., {Casagrande} L., {Asplund} M., 2010, \aap, 524,
  A93

\bibitem[{{Chiavassa} et~al.(2012)}]{chi12}
{Chiavassa} A., {Bigot} L., {Kervella} P., {Matter} A., {Lopez} B., {Collet}
  R., {Magic} Z., {Asplund} M., 2012, \aap, 540, A5

\bibitem[{{Code} et~al.(1976){Code}, {Bless}, {Davis} \& {Brown}}]{code76}
{Code} A.~D., {Bless} R.~C., {Davis} J., {Brown} R.~H., 1976, \apj, 203, 417

\bibitem[{{Cohen} et~al.(1992){Cohen}, {Walker}, {Barlow} \&
  {Deacon}}]{cohen92}
{Cohen} M., {Walker} R.~G., {Barlow} M.~J., {Deacon} J.~R., 1992, \aj, 104,
  1650

\bibitem[{{Cohen} et~al.(2003){Cohen}, {Wheaton} \& {Megeath}}]{cohen03}
{Cohen} M., {Wheaton} W.~A., {Megeath} S.~T., 2003, \aj, 126, 1090

\bibitem[{{Colavita} et~al.(1999)}]{colavita99}
{Colavita} M.~M. et~al., 1999, \apj, 510, 505

\bibitem[{{Cousins}(1980)}]{c80}
{Cousins} A.~W.~J., 1980, South African Astronomical Observatory Circular, 1,
  234

\bibitem[{{Creevey} et~al.(2012)}]{creevey12}
{Creevey} O.~L. et~al., 2012, \aap, 545, A17

\bibitem[{{Cutri} et~al.(2003)}]{cutri03}
{Cutri} R.~M. et~al., 2003, {2MASS All Sky Catalog of point sources.} The IRSA
  2MASS All-Sky Point Source Catalog, NASA/IPAC Infrared Science
  Archive.~http://irsa.ipac.caltech.edu/applications/Gator/

\bibitem[{{Datson} et~al.(2012){Datson}, {Flynn} \& {Portinari}}]{dfp12}
{Datson} J., {Flynn} C., {Portinari} L., 2012, \mnras, 426, 484

\bibitem[{{Datson} et~al.(2014){Datson}, {Flynn} \& {Portinari}}]{datson2013}
{Datson} J., {Flynn} C., {Portinari} L., 2014, astro-ph/1401.1316

\bibitem[{{Davis} \& {Tango}(1986)}]{dt86}
{Davis} J., {Tango} W.~J., 1986, \nat, 323, 234

\bibitem[{{Davis} et~al.(2011){Davis}, {Ireland}, {North}, {Robertson}, {Tango}
  \& {Tuthill}}]{davis11}
{Davis} J., {Ireland} M.~J., {North} J.~R., {Robertson} J.~G., {Tango} W.~J.,
  {Tuthill} P.~G., 2011, \pasa, 28, 58

\bibitem[{{Demory} et~al.(2009)}]{d09}
{Demory} B.~O. et~al., 2009, \aap, 505, 205

\bibitem[{{di Benedetto}(1998)}]{diBene98}
{di Benedetto} G.~P., 1998, \aap, 339, 858

\bibitem[{{Dyck} et~al.(1996){Dyck}, {Benson}, {van Belle} \&
  {Ridgway}}]{dyck96}
{Dyck} H.~M., {Benson} J.~A., {van Belle} G.~T., {Ridgway} S.~T., 1996, \aj,
  111, 1705

\bibitem[{{Feltzing} \& {Bensby}(2008)}]{fb08}
{Feltzing} S., {Bensby} T., 2008, Physica Scripta Volume T, 133, 014031

\bibitem[{{Glass}(1974)}]{g74}
{Glass} I.~S., 1974, Monthly Notes of the Astronomical Society of South Africa,
  33, 53

\bibitem[{{Gonz{\'a}lez Hern{\'a}ndez} \& {Bonifacio}(2009)}]{ghb09}
{Gonz{\'a}lez Hern{\'a}ndez} J.~I., {Bonifacio} P., 2009, \aap, 497, 497

\bibitem[{{Gray} et~al.(2003){Gray}, {Corbally}, {Garrison}, {McFadden} \&
  {Robinson}}]{gc03}
{Gray} R.~O., {Corbally} C.~J., {Garrison} R.~F., {McFadden} M.~T., {Robinson}
  P.~E., 2003, \aj, 126, 2048

\bibitem[{{Hanbury Brown} et~al.(1974){Hanbury Brown}, {Davis} \&
  {Allen}}]{hb74}
{Hanbury Brown} R., {Davis} J., {Allen} L.~R., 1974, \mnras, 167, 121

\bibitem[{{Hekker} et~al.(2009)}]{hekker09}
{Hekker} S. et~al., 2009, \aap, 506, 465

\bibitem[{{Hinkley} et~al.(2010)}]{hink10}
{Hinkley} S. et~al., 2010, \apj, 712, 421

\bibitem[{{H{\o}g} et~al.(2000)}]{hog2000}
{H{\o}g} E. et~al., 2000, \aap, 355, L27

\bibitem[{{Holmberg} et~al.(2007){Holmberg}, {Nordstr{\"o}m} \&
  {Andersen}}]{holmberg07}
{Holmberg} J., {Nordstr{\"o}m} B., {Andersen} J., 2007, \aap, 475, 519

\bibitem[{{Holmberg} et~al.(2009){Holmberg}, {Nordstr{\"o}m} \&
  {Andersen}}]{holmberg09}
{Holmberg} J., {Nordstr{\"o}m} B., {Andersen} J., 2009, \aap, 501, 941

\bibitem[{{Huber} et~al.(2012)}]{huber12}
{Huber} D. et~al., 2012, \apj, 760, 32

\bibitem[{{Ireland} et~al.(2008)}]{chara_pavo}
{Ireland} M.~J. et~al., 2008, in Society of Photo-Optical Instrumentation
  Engineers (SPIE) Conference Series, Vol. 7013

\bibitem[{{Kervella} et~al.(2003{\natexlab{a}}){Kervella}, {Th{\'e}venin},
  {Morel}, {Bord{\'e}} \& {Di Folco}}]{ker03}
{Kervella} P., {Th{\'e}venin} F., {Morel} P., {Bord{\'e}} P., {Di Folco} E.,
  2003{\natexlab{a}}, \aap, 408, 681

\bibitem[{{Kervella} et~al.(2003{\natexlab{b}}){Kervella}, {Th{\'e}venin},
  {S{\'e}gransan}, {Berthomieu}, {Lopez}, {Morel} \& {Provost}}]{kervella03}
{Kervella} P., {Th{\'e}venin} F., {S{\'e}gransan} D., {Berthomieu} G., {Lopez}
  B., {Morel} P., {Provost} J., 2003{\natexlab{b}}, \aap, 404, 1087

\bibitem[{{Kervella} et~al.(2008)}]{km08}
{Kervella} P. et~al., 2008, \aap, 488, 667

\bibitem[{{King} et~al.(2005){King}, {Boesgaard} \& {Schuler}}]{k05}
{King} J.~R., {Boesgaard} A.~M., {Schuler} S.~C., 2005, \aj, 130, 2318

\bibitem[{{Lallement} et~al.(2014){Lallement}, {Vergely}, {Valette},
  {Puspitarini}, {Eyer} \& {Casagrande}}]{lallement13}
{Lallement} R., {Vergely} J.~L., {Valette} B., {Puspitarini} L., {Eyer} L.,
  {Casagrande} L., 2014, \aap, 561, A91

\bibitem[{{Leroy}(1993)}]{leroy93}
{Leroy} J.~L., 1993, \aap, 274, 203

\bibitem[{{Ligi} et~al.(2012)}]{Ligi2012}
{Ligi} R. et~al., 2012, \aap, 545, A5

\bibitem[{{Magic} et~al.(2013)}]{magic13a}
{Magic} Z., {Collet} R., {Asplund} M., {Trampedach} R., {Hayek} W., {Chiavassa}
  A., {Stein} R.~F., {Nordlund} {\AA}., 2013, A\&A, 557, 26

\bibitem[{{Matteucci}(2003)}]{matteucci:book}
{Matteucci} F., 2003, {The Chemical Evolution of the Galaxy}

\bibitem[{{Mel{\'e}ndez} \& {Ram{\'{\i}}rez}(2007)}]{melendez07}
{Mel{\'e}ndez} J., {Ram{\'{\i}}rez} I., 2007, \apjl, 669, L89

\bibitem[{{Mel{\'e}ndez} et~al.(2009){Mel{\'e}ndez}, {Asplund}, {Gustafsson} \&
  {Yong}}]{melendez09:twins}
{Mel{\'e}ndez} J., {Asplund} M., {Gustafsson} B., {Yong} D., 2009, \apjl, 704,
  L66

\bibitem[{{Mel{\'e}ndez} et~al.(2010{\natexlab{a}}){Mel{\'e}ndez},
  {Casagrande}, {Ram{\'{\i}}rez}, {Asplund} \& {Schuster}}]{melendez09:lithium}
{Mel{\'e}ndez} J., {Casagrande} L., {Ram{\'{\i}}rez} I., {Asplund} M.,
  {Schuster} W.~J., 2010{\natexlab{a}}, \aap, 515, L3

\bibitem[{{Mel{\'e}ndez} et~al.(2010{\natexlab{b}}){Mel{\'e}ndez}, {Schuster},
  {Silva}, {Ram{\'{\i}}rez}, {Casagrande} \& {Coelho}}]{melendez10}
{Mel{\'e}ndez} J., {Schuster} W.~J., {Silva} J.~S., {Ram{\'{\i}}rez} I.,
  {Casagrande} L., {Coelho} P., 2010{\natexlab{b}}, \aap, 522, A98

\bibitem[{{Menzies} et~al.(1989){Menzies}, {Cousins}, {Banfield} \&
  {Laing}}]{m89}
{Menzies} J.~W., {Cousins} A.~W.~J., {Banfield} R.~M., {Laing} J.~D., 1989,
  South African Astronomical Observatory Circular, 13, 1

\bibitem[{{Michelson} \& {Pease}(1921)}]{mp21}
{Michelson} A.~A., {Pease} F.~G., 1921, \apj, 53, 249

\bibitem[{{Mozurkewich} et~al.(1991)}]{mozu91}
{Mozurkewich} D. et~al., 1991, \aj, 101, 2207

\bibitem[{{Mozurkewich} et~al.(2003)}]{moz03}
{Mozurkewich} D. et~al., 2003, \aj, 126, 2502

\bibitem[{{Mu{\~n}oz Bermejo} et~al.(2013){Mu{\~n}oz Bermejo}, {Asensio Ramos}
  \& {Allende Prieto}}]{munoz13}
{Mu{\~n}oz Bermejo} J., {Asensio Ramos} A., {Allende Prieto} C., 2013, \aap,
  553, A95

\bibitem[{{Nordgren} et~al.(1999)}]{nordgren99}
{Nordgren} T.~E. et~al., 1999, \aj, 118, 3032

\bibitem[{{Nordlander} et~al.(2012){Nordlander}, {Korn}, {Richard} \&
  {Lind}}]{nor12}
{Nordlander} T., {Korn} A.~J., {Richard} O., {Lind} K., 2012, \apj, 753, 48

\bibitem[{{Nordstr{\"o}m} et~al.(2004)}]{nordstrom04}
{Nordstr{\"o}m} B. et~al., 2004, \aap, 418, 989

\bibitem[{{North} et~al.(2007)}]{north07}
{North} J.~R. et~al., 2007, \mnras, 380, L80

\bibitem[{{North} et~al.(2009)}]{north09}
{North} J.~R. et~al., 2009, \mnras, 393, 245

\bibitem[{{Pagel}(1997)}]{pagel:book}
{Pagel} B.~E.~J., 1997, {Nucleosynthesis and Chemical Evolution of Galaxies}

\bibitem[{{Pease}(1931)}]{p31}
{Pease} F.~G., 1931, Ergebn. Exakten Naturw., 10, 84, 10, 84

\bibitem[{{Pereira} et~al.(2013){Pereira}, {Asplund}, {Collet}, {Thaler},
  {Trampedach} \& {Leenaarts}}]{p13}
{Pereira} T.~M.~D., {Asplund} M., {Collet} R., {Thaler} I., {Trampedach} R.,
  {Leenaarts} J., 2013, \aap, 554, A118

\bibitem[{{Pinsonneault} et~al.(2012){Pinsonneault}, {An},
  {Molenda-{\.Z}akowicz}, {Chaplin}, {Metcalfe} \& {Bruntt}}]{pin12}
{Pinsonneault} M.~H., {An} D., {Molenda-{\.Z}akowicz} J., {Chaplin} W.~J.,
  {Metcalfe} T.~S., {Bruntt} H., 2012, \apjs, 199, 30

\bibitem[{{Porto de Mello} \& {da Silva}(1997)}]{demello97}
{Porto de Mello} G.~F., {da Silva} L., 1997, \apjl, 482, L89+

\bibitem[{{Ram{\'{\i}}rez} \& {Mel{\'e}ndez}(2005{\natexlab{a}})}]{rm05a}
{Ram{\'{\i}}rez} I., {Mel{\'e}ndez} J., 2005{\natexlab{a}}, \apj, 626, 446

\bibitem[{{Ram{\'{\i}}rez} \& {Mel{\'e}ndez}(2005{\natexlab{b}})}]{rm05b}
{Ram{\'{\i}}rez} I., {Mel{\'e}ndez} J., 2005{\natexlab{b}}, \apj, 626, 465

\bibitem[{{Ram{\'{\i}}rez} et~al.(2009){Ram{\'{\i}}rez}, {Mel{\'e}ndez} \&
  {Asplund}}]{ramirez09}
{Ram{\'{\i}}rez} I., {Mel{\'e}ndez} J., {Asplund} M., 2009, \aap, 508, L17

\bibitem[{{Ram{\'{\i}}rez} et~al.(2012)}]{r12}
{Ram{\'{\i}}rez} I. et~al., 2012, \apj, 752, 5

\bibitem[{{Ricker} et~al.(2010)}]{ricker10}
{Ricker} G.~R. et~al., 2010, in American Astronomical Society Meeting Abstracts
  215. Bulletin of the American Astronomical Society, Vol.~42, p. 450.06

\bibitem[{{Rieke} et~al.(2008)}]{rieke08}
{Rieke} G.~H. et~al., 2008, \aj, 135, 2245

\bibitem[{{Ruchti} et~al.(2013){Ruchti}, {Bergemann}, {Serenelli}, {Casagrande}
  \& {Lind}}]{Ruchti2013}
{Ruchti} G.~R., {Bergemann} M., {Serenelli} A., {Casagrande} L., {Lind} K.,
  2013, \mnras, 429, 126

\bibitem[{{Sbordone} et~al.(2010)}]{sbo10}
{Sbordone} L. et~al., 2010, \aap, 522, A26

\bibitem[{{Silva Aguirre} et~al.(2011)}]{vsa11}
{Silva Aguirre} V. et~al., 2011, \apjl, 740, L2

\bibitem[{{Silva Aguirre} et~al.(2012)}]{vsa12}
{Silva Aguirre} V. et~al., 2012, \apj, 757, 99

\bibitem[{{Sousa} et~al.(2011){Sousa}, {Santos}, {Israelian}, {Mayor} \&
  {Udry}}]{sousa11}
{Sousa} S.~G., {Santos} N.~C., {Israelian} G., {Mayor} M., {Udry} S., 2011,
  \aap, 533, A141

\bibitem[{{Stello} et~al.(2009){Stello}, {Chaplin}, {Basu}, {Elsworth} \&
  {Bedding}}]{stello09}
{Stello} D., {Chaplin} W.~J., {Basu} S., {Elsworth} Y., {Bedding} T.~R., 2009,
  \mnras, 400, L80

\bibitem[{{Takeda} et~al.(2009){Takeda}, {Kang}, {Han}, {Lee} \& {Kim}}]{tak09}
{Takeda} Y., {Kang} D.~I., {Han} I., {Lee} B.~C., {Kim} K.~M., 2009, \pasj, 61,
  1165

\bibitem[{{Torres} et~al.(2010){Torres}, {Andersen} \& {Gim{\'e}nez}}]{tor10}
{Torres} G., {Andersen} J., {Gim{\'e}nez} A., 2010, \aapr, 18, 67

\bibitem[{{Valenti} \& {Fischer}(2005)}]{valenti05}
{Valenti} J.~A., {Fischer} D.~A., 2005, \apjs, 159, 141

\bibitem[{{van Belle} \& {von Braun}(2009)}]{vBvB09}
{van Belle} G.~T., {von Braun} K., 2009, \apj, 694, 1085

\bibitem[{{van Belle} et~al.(2007){van Belle}, {Ciardi} \& {Boden}}]{vb07}
{van Belle} G.~T., {Ciardi} D.~R., {Boden} A.~F., 2007, \apj, 657, 1058

\bibitem[{{van Leeuwen}(2007)}]{vanLeeuwen07}
{van Leeuwen} F., 2007, \aap, 474, 653

\bibitem[{{VandenBerg} et~al.(2010){VandenBerg}, {Casagrande} \&
  {Stetson}}]{vandenberg10}
{VandenBerg} D.~A., {Casagrande} L., {Stetson} P.~B., 2010, \aj, 140, 1020

\bibitem[{{von Braun} et~al.(2011)}]{vB11}
{von Braun} K. et~al., 2011, \apj, 740, 49

\bibitem[{{White} et~al.(2013)}]{w13}
{White} T.~R. et~al., 2013, \mnras, 433, 1262

\bibitem[{{Wielen} et~al.(1996){Wielen}, {Fuchs} \& {Dettbarn}}]{wfd96}
{Wielen} R., {Fuchs} B., {Dettbarn} C., 1996, \aap, 314, 438

\end{thebibliography}

\end{document}